\documentclass[12pt,preprint]{aastex}
\usepackage{epsfig}
\usepackage{natbib}
\usepackage{graphicx}
\usepackage{mathrsfs,amssymb}
\usepackage{amsmath}
\newcommand	\beq	{\begin{equation}}	%{\begin{displaymath}}
\newcommand	\eeq	{\end{equation}}	%{\end{displaymath}}
\newcommand       \Angstrom     {\,{\rm \AA}}

\newcommand       \cm           {\,{\rm cm}}

\newcommand       \g            {\,{\rm g}}
\newcommand       \K            {\,{\rm K}}

\newcommand       \nH           {n_{\rm H}}
\newcommand       \NH           {N_{\rm H}}
\newcommand       \simlt        {\lesssim}

\newcommand       \um           {\mu{\rm m}}
\newcommand       \mum          {\,{\rm \mu m}}
\newcommand       \ppm          {\,{\rm ppm}}

\newcommand       \mH           {m_{\rm H}}
\newcommand       \MH           {M_{\rm H}}
\newcommand       \simali       {\sim\,}
\newcommand       \magni        {\,{\rm mag}}
\newcommand       \rmH          {{\rm H}}
\newcommand       \HH           {\,{\rm H}}
\newcommand       \Msil           {M_{\rm sil}}

\newcommand       \rhosil        {\rho_{\rm sil}}
\newcommand       \rhogra        {\rho_{\rm gra}}
\newcommand       \rhoC          {\rho_{\rm C}}

\newcommand	  \xism         {\left[{\rm X/H}\right]_{\rm ISM}}
\newcommand	  \csun         {\left[{\rm C/H}\right]_{\odot}}

\newcommand	  \fesun        {\left[{\rm Fe/H}\right]_{\odot}}
\newcommand	  \mgsun        {\left[{\rm Mg/H}\right]_{\odot}}
\newcommand	  \sisun        {\left[{\rm Si/H}\right]_{\odot}}

\newcommand	  \cstar         {\left[{\rm C/H}\right]_{\star}}

\newcommand	  \festar        {\left[{\rm Fe/H}\right]_{\star}}
\newcommand	  \mgstar        {\left[{\rm Mg/H}\right]_{\star}}
\newcommand	  \sistar        {\left[{\rm Si/H}\right]_{\star}}
\newcommand	  \cism         {\left[{\rm C/H}\right]_{\rm ISM}}

\newcommand	  \siism        {\left[{\rm Si/H}\right]_{\rm ISM}}
\newcommand	  \xdust        {\left[{\rm X/H}\right]_{\rm dust}}
\newcommand	  \cdust        {\left[{\rm C/H}\right]_{\rm dust}}

\newcommand	  \fedust       {\left[{\rm Fe/H}\right]_{\rm dust}}
\newcommand	  \mgdust       {\left[{\rm Mg/H}\right]_{\rm dust}}
\newcommand	  \sidust       {\left[{\rm Si/H}\right]_{\rm dust}}
\newcommand	  \xgas         {\left[{\rm X/H}\right]_{\rm gas}}
\newcommand	  \cgas        {\left[{\rm C/H}\right]_{\rm gas}}

\newcommand	  \mux         {\mu_{\rm X}}

\newcommand	  \musi        {\mu_{\rm Si}}
\newcommand	  \mufe        {\mu_{\rm Fe}}
\newcommand	  \mumg        {\mu_{\rm Mg}}

\newcommand	  \Vdust        {V_{\rm dust}}
\newcommand	  \Vsil         {V_{\rm sil}}

\newcommand	  \VC           {V_{\rm C}}
\newcommand	  \Fsil         {F_{\rm sil}}

\newcommand	  \FC           {F_{\rm C}}
\newcommand	  \Aint         {A_{\rm int}}
\newcommand	  \Aintp        {A_{\rm int}^{\prime}}
\newcommand       \Bsil          {B_{\rm sil}}
\newcommand       \Bgra          {B_{\rm gra}}

\def    \acC		{a_{c,{\rm C}}}
\def    \acS		{a_{c,{\rm S}}}

\def    \alphaC		{\alpha_{\rm C}}
\def    \alphaS		{\alpha_{\rm S}}

\def    \rhoC		{\rho_{\rm C}}
\def    \Bsil		{B_{\rm S}}
\def    \Bgra		{B_{\rm C}}

\def    \xo             {x_{\rm o}}
\newcommand{\figwidth}{4.0in}
\countdef\decade=200
\advance\decade by \year
\countdef\hours=201
\advance\hours by \time
\divide\hours by 60
\countdef\mins=202
\advance\mins by \hours
\multiply\mins by 60
\multiply\hours by 100
\countdef\miltime=203
\advance\miltime by \hours
\advance\miltime by \time
\advance\miltime by -\mins

\shorttitle{Carbon and the Ultraviolet Extinction}
\title{
Probing the Role of Carbon 
in the Ultraviolet Interstellar Extinction
}
\author{Ajay Mishra and Aigen Li}
\affil{Department of Physics and Astronomy,
        University of Missouri,
        Columbia, MO 65211, USA;
        {\sf amishra@mail.missouri.edu,
             lia@missouri.edu}}

\begin{document}

\begin{abstract}
We probe the role of carbon in the ultraviolet (UV)
extinction by examining the relations between the amount 
of carbon required to be locked up in dust $\cdust$ with 
the 2175$\Angstrom$ extinction bump and the far-UV
extinction rise, based on an analysis of the extinction 
curves along 16 Galactic sightlines for which 
the gas-phase carbon abundance is known 
and the 2175$\Angstrom$ extinction bump 
exhibits variable strengths and widths.
We derive $\cdust$ from the %model-independent 
Kramers-Kronig relation which relates 
the wavelength-integrated extinction
to the total dust volume.
This approach is less model-dependent
since it does not require the knowledge 
of the detailed optical properties and
size distribution of the dust. 
We also derive $\cdust$ from fitting the observed
UV/optical/near-infrared extinction
with a mixture of amorphous silicate and graphite.
We find that the carbon depletion $\cdust$ 
tends to correlate with the strength of 
the 2175$\Angstrom$ bump, 
while the abundance of silicon depleted in dust 
shows no correlation with the 2175$\Angstrom$ bump. 
This supports graphite 
or polycyclic aromatic hydrocarbon (PAH) molecules 
as the possible carrier of the 2175$\Angstrom$ bump. 
We also see that $\cdust$ shows 
a trend of correlating with $1/R_V$,
where $R_V$ is the total-to-selective extinction ratio, 
suggesting that the far-UV extinction is more likely 
produced by small carbon dust than by small silicate dust.
\end{abstract}

\keywords{dust, extinction --- ISM: abundances  --- ISM: clouds}

\section{Introduction}\label{sec:intro}
Dust is a ubiquitous feature of the cosmos. 
It has dramatic effects on the interstellar medium (ISM)
by affecting the interstellar conditions, 
participating in the physical and chemical processes 
occuring in the ISM, and shaping the appearance 
of dusty objects as an absorber, scatterer, 
and emitter of electromagnetic radiation,
and as an efficient catalyst 
for the formation of H$_2$
and an agent for photoelectrically 
heating the interstellar gas. 
To understand the interaction of 
the solid dust particles 
in the space between stars 
(i.e., interstellar dust) with radiation
and the dust-related interstellar processes,
the knowledge of the dust composition and size
is required.

What is interstellar dust made of?
The ubiquitous detection of the 9.7$\mum$ Si--O 
stretching and 18$\mum$ O--Si--O bending absorption 
features in interstellar regions implies that
silicate must be a major component (Henning 2010).
These features are broad, smooth and featureless,
suggesting that interstellar silicate is amorphous
(Li \& Draine 2001a, Kemper et al.\ 2004, Li et al.\ 2008). 
Another absorption feature ubiquitously seen in
the diffuse ISM of the Milky Way and external galaxies
is the 3.4$\mum$ feature, commonly attributed to 
the C--H stretch in some sorts of aliphatic hydrocarbon
materials (see Pendleton \& Allamandola 2002).
Depending on their H fractions and the preparation method,
the experimentally-synthesized aliphatic hydrocarbon
analogues show considerable variation in the C--H band 
strength (e.g., see Furton, Laiho, \& Witt 1999). 
Therefore, it has not been possible to 
use the measured 3.4$\mum$ feature to determine
the abundance of aliphatic hydrocarbon material
in the ISM (e.g., see Chiar et al.\ 2013).

One may infer the amount of silicate dust $\Msil$
%(relative to the hydrogen mass $\MH$)
from the interstellar Si, Mg, and Fe abundances.
Let $\xism$ be the total interstellar abundance
of element X relative to H,
$\xgas$ be the amount of X in the gas phase,
and $\xdust$ be the amount of X contained in dust.
As an element will be in gas-phase 
and/or locked up in solid-phase,  
we will obviously have $\xdust$\,=\,$\xism-\xgas$.
Let $\mux$ be the atomic weight of X
($\mux$\,=\,12, 16, 56, 24, and 28
for carbon, oxygen, iron, magnesium, and silicon).
If we assume a stoichiometric composition of
Mg$_{\rm 2x}$Fe$_{\rm 2(1-x)}$SiO$_4$
for silicate (i.e., each silicon atom
corresponds to four oxygen atoms), 
the interstellar silicate dust mass 
relative to hydrogen is 
\begin{equation}\label{eq:Msil2MH}
\Msil/\MH = \mufe\fedust + \mumg\mgdust
          + \musi\sidust + 4\times \mu_{\rm O}\sidust ~~.
\end{equation}
The exact interstellar abundances for the dust-forming 
elements are unknown (see Snow \& Witt 1996, Sofia 2004,
Li 2005, Jenkins 2009). 
The abundances of the Sun, the proto-Sun,
unevolved early B stars, and F and G stars
are often adopted as the interstellar reference
abundance standards.
If we adopt the proto-Sun abundances of
$\fesun=34.7\pm2.4\ppm$,
$\mgsun=41.7\pm1.9\ppm$, and
$\sisun=40.7\pm1.9\ppm$ (Lodders 2003)  
for the interstellar reference abundances
and assume that Fe, Mg and Si are essentially
completely locked up in dust
(i.e., $\xdust=\xism$ for Fe, Mg and Si),
we obtain $\Msil/\MH \approx 6.69\times10^{-3}$.
Let $C_{\rm abs}(9.7\mu {\rm m})/V
=1.00\times10^{4}\cm^{-1}$ 
be the peak absorption cross section per volume 
of silicate dust at 9.7$\mum$ (Draine \& Lee 1984)
and $A_V/\NH = 5.3\times10^{-22}\magni\cm^2\HH^{-1}$ 
be the visual extinction $A_V$ 
per hydrogen column $\NH$ (Whittet 2003). 
One may estimate the ratio of the optical depth 
of the 9.7$\mum$ silicate absorption feature to 
$A_V$ to be 
\beq
\Delta\tau_{9.7}/A_V\approx 
\left[C_{\rm abs}(9.7\mu {\rm m})/V\right]/\rho_{\rm sil} 
\times \left(\Msil/\MH\right)\times 
\left(\NH/A_V\right)\times\mH\approx 1/16.7 ~~,
\eeq
where $\rho_{\rm sil}\approx3.5\g\cm^{-3}$
is the mass density of silicate and 
$\mH=1.66\times10^{-24}\g$
is the mass of a hydrogen atom.
This is close to the mean value of 
$\Delta\tau_{9.7}/A_V\approx1/18.5$ 
observed for the local diffuse ISM
(see Draine 2003a), 
suggesting that the proto-Sun Fe, Mg and Si 
abundances are sufficient 
to account for the observed 9.7$\mum$ silicate 
absorption feature.\footnote{%
   \label{ftnt:abund}
   If we adopt the abundances of
   unevolved early B-type stars
   ($\festar=27.5\pm2.5\ppm$,
    $\mgstar=36.3\pm4.2\ppm$, and
    $\sistar=31.6\pm1.5\ppm$; 
    Przybilla et al.\ 2008) 
   for the interstellar reference abundances,
   we obtain $\Delta\tau_{9.7}/A_V\approx1/21.0$.
   This is somewhat lower than the observed
   value of $\Delta\tau_{9.7}/A_V\approx1/18.5$.  
   The abundances of the silicate-forming elements
   of the Sun
   ($\fesun=31.6\pm2.9\ppm$,
    $\mgsun=39.8\pm3.7\ppm$, and
    $\sisun=32.4\pm2.2\ppm$; 
    Asplund et al.\ 2009)
    and of young F and G stars
    ($\festar=28.2\pm7.8\ppm$,
     $\mgstar=42.7\pm16.7\ppm$, and
     $\sistar=39.8\pm12.8\ppm$; 
     Sofia \& Meyer 2001)
    are somewhat intermediate 
    between that of the proto-Sun
    and early B stars,
    except the magnesium abundance
    of young F and G stars 
    is a little bit higher than that 
    of the proto-Sun and early B stars
    (but we note that the abundances of
    young F and G stars have large uncertainties;
    see Sofia \& Meyer 2001).
   %
   % C/H = 354.8 <-- young F and G stars; Sofia \& Meyer; 
   % O/H = 446.7 <-- young F and G stars; 
   %
   % C/H = 209\pm 15 <-- B stars; Przybilla, Nieva;
   % O/H = 575\pm41 <-- young F and G stars;
   %
   % C/H = 224pm <-- Sun; Asplund
   % O/H =  <-- Sun;
   %
   % C/H = 288ppm <-- proto-Sun; Lodders;
   % O/H =  <-- proto-Sun;
   }

However, silicate alone is not sufficient to account 
for the observed interstellar extinction.
As shown by Purcell (1969), the Kramers-Kronig relation 
can be used to 
%place a lower limit on
%the total amount of dust in the ISM.
constrain the dust quantity in the ISM.
Let $A_\lambda$ be the extinction at wavelength $\lambda$ 
and $A_\lambda/\NH$ be the extinction per H nucleon.
As shown in Li (2005), 
the Kramers-Kronig relation %can be used to 
relates the wavelength-integrated extinction
($\int_{0}^{\infty} A_\lambda/\NH\,d\lambda$)
to the total volume occupied by dust per H nucleon 
($\Vdust/{\rm H}$) through 
\begin{equation}\label{eq:kk}
\int_{0}^{\infty} \frac{A_\lambda}{\NH} d\lambda 
= 1.086\times 3 \pi^2 F \frac{\Vdust}{\rm H} ~~,
\end{equation}
%where the dimensionless factor $F$ is the orientationally-averaged 
%polarizability relative to the polarizability of an equal-volume 
%conducting sphere, depending only upon the grain shape and 
%the static (zero-frequency) dielectric constant 
%$\varepsilon_0$ of the grain material (Purcell 1969).
where $F$ is a dimensionless factor 
and depends only upon the grain shape and 
the static (zero-frequency) dielectric constant 
$\varepsilon_0$ of the grain material (see below). 
%(Purcell 1969).
%
The interstellar extinction per H nucleon 
($A_\lambda/\NH$) 
is not known for all wavelengths,
but for a limited range of wavelengths. 
We approximate $\int_{0}^{\infty} A_\lambda/\NH\,d\lambda$
by $\int_{912\Angstrom}^{1000\mum} A_\lambda/\NH\,d\lambda$,
with the latter derived from 
the silicate-graphite-PAH model of 
Weingartner \& Draine (2001; hereafter WD01) 
which closely reproduces the observed extinction
from the far-UV to the near-IR
and up to $\lambda\approx30\mum$
(see Figure~16 of Li \& Draine 2001b). 
The WD01 model gives a reasonably accurate 
estimate for the extinction at $\lambda>30\mum$ 
which is observationally unknown
since this model closely fits 
the observed dust thermal emission
from the near-IR to submillimeter.\footnote{%
  The WD01 model reproduces the observed 
  UV/optical extinction 
  as well as the albedo 
  (see Figure~15 of Li \& Draine 2001b).
  This implies that the WD01 model 
  gives a reasonably accurate description
  on how interstellar dust {\it absorbs} starlight. 
  %...of the total power absorbed by interstellar dust,
  %and by implication, the total dust-emitted power. 
  On the other hand,
  the fact that the WD01 model closely 
  reproduces the observed IR emission
  indicates that this model also gives a reasonably 
  accurate description on how interstellar dust 
  {\it emits} in the IR and of its IR opacities.   
  }
For the diffuse ISM, 
%%the mean extinction per H nucleon $A_\lambda/\NH$ 
%%is fairly well-determined from 
%%the far UV to the infrared (IR) and 
%the integration of the mean extinction 
%over the wavelength range 
%$912\Angstrom \le \lambda \le 1000\mum$ 
%is approximately 
the WD01 model gives
$\int_{912\Angstrom}^{1000\mum} 
A_\lambda/\NH\,d\lambda
\approx 1.49\times 
10^{-25}\magni\cm^{3}\HH^{-1}$.\footnote{%
  In the far-IR, the extinction rapidly
  declines with wavelength
  (e.g., $A_\lambda\propto\lambda^{-2}$,
   see Table~6 of Li \& Draine 2001b)
   and therefore, the contribution 
   of the extinction at $\lambda>1000\mum$
   to the extinction integration is
   negligible. Indeed,   
   the wavelength-integrated extinction 
   $\int A_\lambda/\NH\,d\lambda$
   only increases $\simali$1.6\% by extending
   the upper integral limit
   from $\lambda=1000\mum$ to 10,000$\mum$.  
  }
Since $A_\lambda$ is a positive number 
and $\Vdust/\NH$ is obtained by integrating
$A_\lambda/\NH$ over all wavelengths
(see eq.\,\ref{eq:kk}),
therefore, the integration of $A_\lambda/\NH$
over a {\it finite} wavelength range 
can be used to 
%obtain a {\it lower} bound on $V_{\rm dust}/\rmH$.
place a {\it lower} limit on
the total amount of dust in the ISM.

If we take the dust to be spheroids
with semiaxes $a,b,b$ 
(prolate if $a/b>1$, oblate if $a/b<1$),
then $F$ is related to the static dielectric 
constant $\varepsilon_0$ 
and the ``depolarization factors'' 
$L_a$ and $L_b=(1-L_a)/2$ through
(Draine 2003b)
\beq
F(a/b,\varepsilon_0) \equiv \frac{(\varepsilon_0-1)}{9}
\left[
\frac{1}{(\varepsilon_0-1)L_a + 1} + \frac{2}{(\varepsilon_0-1)L_b + 1}
\right]
~~~,
\label{eq:Fdef}
\eeq
where for prolates 
\beq
L_a = \frac{1-e^2}{e^2}
\left[\frac{1}{2e}\ln\left(\frac{1+e}{1-e}\right)-1\right]
~~~,
\eeq
and for oblates 
\beq
L_a = \frac{1+e^2}{e^2}\left[1-\frac{1}{e}\arctan(e)\right]
~~~,
\eeq
where
\beq
e\equiv |1-(b/a)^2|^{1/2} ~~~.
\eeq

In Figure~\ref{fig:dpol} we show the $F$ factors 
for silicate, Fe$_2$O$_3$, amorphous carbon, and
graphite. These dust species are often invoked to
account for the interstellar extinction. 
They also represent materials from being dielectric
(silicate with $\varepsilon_0\approx6.4$,\footnote{%
   Davis et al.\ (1988) measured the temperature 
   dependence of the static dielectric constant 
   $\varepsilon_0$ of naturally occurring monocrystalline 
   forsterite Mg$_{1.86}$Fe$_{0.14}$SiO$_4$.
   %The measurements were made over a frequency range
   %of 11\,kHz to 13\,MHz.
   %They found $\varepsilon_0$ depends linearly upon
   %the temperature in the range 77\,K to 383\,K. 
   They found that $\varepsilon_0$ depends linearly upon
   the temperature: 
   $\varepsilon_0 \approx 6.3 + 0.0064\times T$.
   For typical dust temperatures of $T\approx20\K$
   in the ISM (Li \& Draine 2001b), 
   one obtains $\varepsilon_0\approx6.4$ for silicates.
   }
Fe$_2$O$_3$ with $\varepsilon_0\approx16$) 
to somewhat conducting 
(amorphous carbon with $\varepsilon_0\approx160$) 
to highly conducting 
(graphite with $\varepsilon_0\rightarrow\infty$). 
For silicate spheres we have $F\approx0.64$.
%and $F\approx 1.00$ for graphite spheres.
For modestly elongated 
or flattened silicate spheroids,
$F$ does not exceed unity:
$F\approx0.73$ for $a/b=3$ 
(Greenberg \& Li 1996)\footnote{%
  Greenberg \& Li (1996) modeled 
  the 9.7 and 18$\mum$ silicate polarization 
  features toward the Becklin-Neugebauer (BN) object
  in terms of core-mantle spheroids.   
  They found that the $a/b=3$ prolate shape 
  provides the best fit.
  } 
and $F\approx0.68$ for $a/b=1/2$ 
(Lee \& Draine 1985).\footnote{%
  The $a/b=1/2$ oblate shape was shown to 
  provide a better match than any other shapes 
  to the 3.1$\mum$ ice polarization feature 
  (Lee \& Draine 1985) and the BN 9.7$\mum$ silicate 
  polarization (Hildebrand \& Dragovan 1995). 
  Unlike Greenberg \& Li (1996),
  Hildebrand \& Dragovan (1995) considered bare 
  silicate dust and only modeled the 9.7$\mum$
  polarization.
  }
%Even for conducting graphite spheroids,
%$F$ does not deviate much from unity
%unless they are extremely elongated or flattened:
%$F\approx1.51$ for $a/b=3$ and
%$F\approx1.15$ for $a/b=1/2$.
Assuming the proto-Sun abundances for 
the interstellar reference abundances,
the total silicate volume per H nucleon is 
\beq
\Vsil/\rmH\approx \left(\Msil/\MH\right)/\rho_{\rm sil}
\times \mH\approx3.17\times10^{-27}\cm^3\HH^{-1} ~~.
\eeq 
With $F\approx0.7$ for moderately elongated
or flattened silicate spheroids, 
the wavelength-integrated extinction 
arising from the silicate dust component is at most 
$\approx7.14\times10^{-26}\magni\cm^3\HH^{-1}$.\footnote{%
  This reduces to 
  $\approx5.68\times10^{-26}\magni\cm^3\HH^{-1}$
  if the Fe, Mg and Si abundances of
  unevolved early B stars 
  (Przybilla et al.\ 2008) are adopted.
  The $\Vsil/\rmH$ values obtained with
  the abundances of young F and G stars
  (Sofia \& Meyer 2001) 
  or the Sun (Asplund et al.\ 2009)
  are intermediate between that of the proto-Sun
  and that of the early B stars.
  }
Apparently, this is not sufficient to account for 
the observed lower limit of 
$\approx1.49\times10^{-25}\magni\cm^3\HH^{-1}$.
Therefore, there must exist other dust component(s),
with carbon dust being the most popular one.\footnote{%
  Other dust species such as iron and iron oxides 
  may also be present in the ISM. 
  However, if a substantially large fraction
  of the interstellar Fe/H abundance is consumed 
  by iron or iron oxides, the interstellar silicate
  grains will not be ``dirty'' enough to 
  account for the observed optical/near-IR 
  extinction (see Draine \& Lee 1984).
  Moreover, iron oxides have absorption peaks
  at $\simali$9.2, 18, 21 and 30$\mum$
  for hematite $\alpha$-Fe$_2$O$_3$
  and $\simali$17 and 25$\mum$
  for magnetite Fe$_3$O$_4$
  which are not seen in the ISM
  (see Zhang, Jiang, \& Li 2009).
  }
As a matter of fact, all modern dust models 
invoke some form of carbonaceous materials
including graphite 
(Mathis et al.\ 1977, hereafter MRN; Draine \& Lee 1984), 
``organic refractory'' (Li \& Greenberg 1997),
and hydrogenated amorphous carbon 
(HAC; Duley et al.\ 1989, Jones et al.\ 2013). 
We note that carbon-bearing dust species 
have revealed their presence in the ISM 
%by the ubiquitous detection of 
through the 3.4$\mum$ aliphatic 
C--H absorption feature 
%in the diffuse ISM 
(Pendleton \& Allamandola 2002),
the ``unidentified'' infrared emission
(UIE) bands at 3.3, 6.2, 7.7, 8.6, 11.3 
and 12.7$\mum$ (L\'eger \& Puget 1984, 
Allamandola et al.\ 1985),
and the ``extended Red emission'' (ERE)
attributed to the photoluminescence 
of some form of carbon dust 
(Witt \& Vijh 2004).

What role does the carbonaceous dust component play
in the UV/optical extinction? as elaborated above,
it must account for $\simali$52\% of the total interstellar
extinction. Spectroscopically, the 2175$\Angstrom$ extinction 
bump, the strongest absorption feature observed in the ISM,
was widely attributed to small graphite grains
(Stecher \& Donn 1965, Draine \& Malhotra 1993, Mathis 1994)
or polycyclic aromatic hydrocarbon (PAH) molecules
(Joblin et al.\ 1992, Li \& Draine 2001b,
Cecchi-Pestellini et al.\ 2008, 
Steglich et al.\ 2010, Mulas et al.\ 2013).
It was thought that the $\pi$--$\pi^{\ast}$ transition 
in $sp^{2}$ hybridization of aromatic carbon causes
the 2175$\Angstrom$ extinction bump.
However, Parvathi et al.\ (2012) recently examined
the correlation of the carbon depletion $\cdust$ 
(i.e., the number of carbon atoms [relative to H]
locked up in dust) with the strength of the 2175$\Angstrom$
extinction bump for 21 Galactic lines of sight 
and found no correlation. 
This provides a severe challenge to the hypothesis
of graphite or PAHs as the carriers of 
the 2175$\Angstrom$ bump.

Parvathi et al.\ (2012) derived the dust-phase 
carbon abundance $\cdust$ for each sightline
by subtracting the gas-phase carbon abundance $\cgas$
from the interstellar reference abundance $\cism$:
$\cdust=\cism-\cgas$.
The gas-phase carbon abundance $\cgas$ 
was determined by Sofia et al.\ (2011) 
for six translucent sightlines 
and by Parvathi et al.\ (2012) for 15 sightlines.
Their analysis was based on the strong 1334$\Angstrom$ 
absorption line of CII observed by 
the {\it Space Telescope Imaging Spectrograph} 
(STIS) on board the {\it Hubble Space Telescope} (HST).
As described in Parvathi et al.\ (2012),
these sightlines span a variety of Galactic disk 
environments which are characterized by different 
extinction and sample paths ranging over three orders 
of magnitude in $\langle\nH\rangle$,
the average density of hydrogen.
They have all been observed with 
the {\it Far-Ultraviolet Spectroscopic Explorer} (FUSE) 
and had their total neutral hydrogen 
(i.e., HI and H$_2$) column densities measured 
by Cartledge et al.\ (2004, 2006).
Parvathi et al.\ (2012) took the total interstellar 
carbon abundance to be $\cism = 464\ppm$.
They assumed that the gas-phase carbon abundance 
toward HD\,206773 ($\cgas\approx464\pm57\ppm$) 
represents the lower limit for the interstellar 
cosmic carbon abundance.

%What is $\cism$, the total carbon abundance in the ISM? 
%We need a reference carbon abundance to substitute for $\cism$. 
%Like $\siism$, traditional reference abundances 
%used in the ISM studies are the elemental abundances 
%of the Sun (Asplund et al.\ 2009), 
%the proto-Sun (Lodders 2003),  
%and young F and G stars (Sofia \& Meyer 2001).
%
%However, for several sightlines the gas-phase carbon 
%abundances are higher than the reference abundances.

%However, all of these targets, 
%except HD 37903 and HD 206773, 
%also have 1$\sgma$ error bars that will bring 
%them within the solar carbon abundance. 
%
%Interestingly, the sight line toward HD 206773 
%is found to be significantly reddened 
%($E(B-V)\approx0.45$) 
%despite its large gas-phase carbon abundance. 
%We note that the 1334$\Angstrom$ feature 
%toward this star has weak components in 
%both the short- and long-wavelength damping wings 
%of the absorption. 
%
%As mentioned above, absorption superimposed 
%on the main damping wings increases the systematic errors 
%(not included in the error bars), which likely accounts 
%for the high measured abundance here. 
%
%HD 37903 has its CII absorption spread over a wide range 
%of heliocentric velocities, which will make the systematic 
%uncertainties large in the carbon abundance determination 
%in this sight line. Finally, we note that the systematic 
%uncertainties are also likely high toward HD 210809, 
%where the absorption is spread over a wide velocity range.
%
%

In this work we revisit the Galactic sightlines 
considered by Parvathi et al.\ (2012) to probe the role 
of the carbonaceous dust component in the UV extinction, 
with special attention paid to the 2175$\Angstrom$ extinction 
bump. We take two approaches both of which differ from 
that of Parvathi et al.\ (2012). 
One approach is independent of any dust models:
we use the Kramers-Kronig relation of Purcell (1969) 
to relate the wavelength-integrated extinction 
to the total dust volume (see eq.\,\ref{eq:kk})
to infer the carbon depletion $\cdust$ 
for each sightline (see \S\ref{sec:kk}).
We then examine the relation between $\cdust$ and
the 2175$\Angstrom$ extinction bump strength 
and the far-UV extinction rise (see \S\ref{sec:kk}).
Another approach is to model 
the observed extinction curve 
of each sightline in terms of 
the silicate-graphite model (see \S\ref{sec:extmod}).
Again, the required carbon depletion in graphite 
is compared with the 2175$\Angstrom$ extinction 
bump strength and the far-UV extinction rise 
(see \S\ref{sec:extmod}).
The results are discussed in \S\ref{sec:discussion}
and the major conclusions are summarized 
in \S\ref{sec:conclusion}.

% ... we take one step further to 
% understand the possible carrier of 
% the 2175$\Angstrom$ extinction bump ...
% ... between 0.3 to 8$\mum^{-1}$ for 16 sources 
% towards Galactic line of sight to 
% understand the possible carrier associated 
% with the 2175$\Angstrom$ bump. 
%
% These selected sources has well observed 
% extinction curve, well known hydrogen column 
% density $\Hcol$ and gas phase $\ctohg$ abundances. 

% We explore the relation between the strength of 
% the 2175$\Angstrom$ bump and the abundance of 
% carbon and silicon dust to understand 
% the carrier of the bump. 
% We aslo examine the correlations 
% between C or Si depletion with $\Rvinv$, 
% C or Si depletion with far-UV strength, 
% and the 2175$\Angstrom$ bump strength 
% with steepness of far-UV extinction. 

\section{Carbon Depletion Inferred from 
         the Kramers-Kronig Relation\label{sec:kk}}
Among the 15 Galactic sightlines studied by 
Parvathi et al.\ (2012) and the six sightlines
studied by Sofia et al.\ (2011), we select 16 sightlines
for which the extinction curves, hydrogen column densities,
and gas-phase carbon abundances have been well determined.
In Figures~\ref{fig:extmod1},\,\ref{fig:extmod2} we show
the UV/optical/near-IR extinction curves 
of these 16 sources constructed as following,
with the extinction parameters 
taken from Valencic et al.\ (2004)
and Gordon et al.\ (2009). 

The UV extinction at $3.3 < \lambda^{-1} < 8.7\mum^{-1}$
is derived from the following analytical formula
which consists of three parts: 
a linear background,
a Drude profile for the 2175$\Angstrom$ extinction bump,
and a far-UV nonlinear rise 
at $\lambda^{-1} > 5.9\mum^{-1}$
(see Valencic et al.\ 2004, Gordon et al.\ 2009):  
\begin{equation}\label{eq:A2AV}
A_\lambda/A_V = c_1^{\prime} + c_2^{\prime}\,x 
              + c_3^{\prime}\,D(x,\gamma,\xo) 
              + c_4^{\prime}\,F(x) ~~~,
\end{equation}
\begin{equation}
D(x,\gamma,\xo) = \frac{x^2}
  {\left(x^2-\xo^2\right)^2 + x^2\gamma^2} ~~~,
\end{equation}
\begin{equation}
F(x) = \left\{\begin{array}{lr} 
0 ~, & x < 5.9\mum^{-1} ~~~,\\

0.5392\,\left(x-5.9\right)^2 
     + 0.05644\,\left(x-5.9\right)^3 ~, 
 & x \ge 5.9\mum^{-1} ~~~,\\
\end{array}\right.
\end{equation}
where $x\equiv 1/\lambda$ 
is the inverse wavelength in $\mu$m$^{-1}$, 
$c_1^{\prime}$ and $c_2^{\prime}$ define 
the linear background, 
$c_3^{\prime}$ defines the strength of 
the 2175$\Angstrom$ extinction bump
which is approximated by $D(x,\gamma,\xo)$,
a Drude function which peaks at 
$\xo\approx4.6\mum^{-1}$ and has 
a FWHM of $\gamma$, 
and $c_4^{\prime}$ defines 
the nonlinear far-UV rise.
This parametrization was originally introduced
by Fitzpatrick \& Massa (1990; hereafter FM90)
for the interstellar reddening
\begin{equation}
E(\lambda-V)/E(B-V) = R_V \left(A_\lambda/A_V - 1\right)
              = c_1 + c_2\,x 
              + c_3\,D(x,\gamma,\xo) 
              + c_4\,F(x) ~~~,
\end{equation}
where $E(\lambda-V) = A_\lambda - A_V$,
$E(B-V) = A_B - A_V$, 
and $A_B$ is the $B$-band extinction.
It is easy to find the relationship
between the FM90 parameters and that
of Valencic et al.\ (2004)
and Gordon et al.\ (2009):  
\begin{equation}
c_j^{\prime} = \left\{\begin{array}{lr} 
c_j/R_V + 1 ~, & j=1 ~~~,\\
c_j/R_V ~, & j=2, 3, 4 ~~~,\\
\end{array}\right.
\end{equation}
where $R_V\equiv A_V/E(B-V)$ is
the total-to-selective extinction ratio.
The $c_j^{\prime}$ parameters of 
the selected 16 sources are taken 
from Valencic et al.\ (2004)
and Gordon et al.\ (2009). 
Finally,
for a given $R_V$,
the optical/near-IR extinction 
at $0.3 < \lambda^{-1} < 3.3\mum^{-1}$
is computed from the CCM parametrization 
(see Cardelli et al.\ 1989).
We then smoothly join the UV extinction
at $\lambda^{-1} > 3.3\mum^{-1}$
to the optical/near-IR extinction. 

\begin{figure}[ht]
\begin{center}
\epsfig{
        file=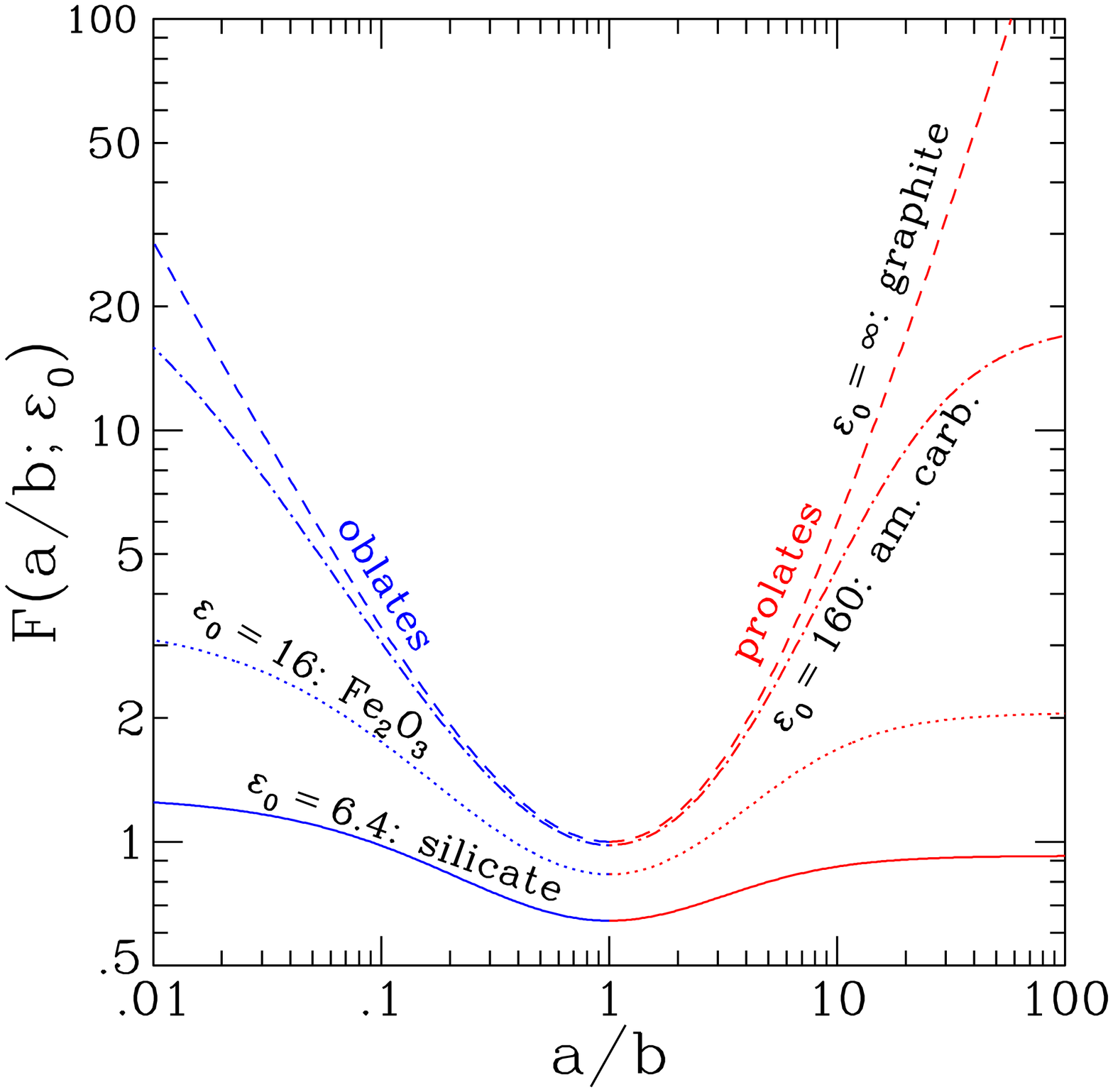,
        width=\figwidth,angle=0}
\end{center}\vspace*{-1em}
\caption{
        \label{fig:dpol}
        \footnotesize
        The $F(\varepsilon_0;{\rm shape})$ factor
        as a function of the axial ratio $a/b$
        for silicate (solid line), 
        Fe$_2$O$_3$ (dotted line),
        amorphous carbon (dot-dashed line), 
        and graphite (dashed line)
        of which the static dielectric constants
        are approximately 
        $\varepsilon_0$\,$\approx$\,6.4, 16, 160 
        and $\infty$ (Li 2005).
        The grains are taken to be spheroidal
        with $a$ and $b$ being the semiaxis 
        along and perpendicular to the symmetry axis
        of the spheroid, respectively.
        The blue lines are for oblates ($a/b<1$) 
        and the red lines are for prolates ($a/b>1$).
        For modestly elongated ($a/b\simlt 3$) 
        or flattened ($b/a\simlt 3$) silicate dust,
        the $F$ factor is always smaller than unity. 
        }
\end{figure}

\begin{figure*}
\centering	
\vspace{-5mm}
\includegraphics[width=0.80\textwidth]{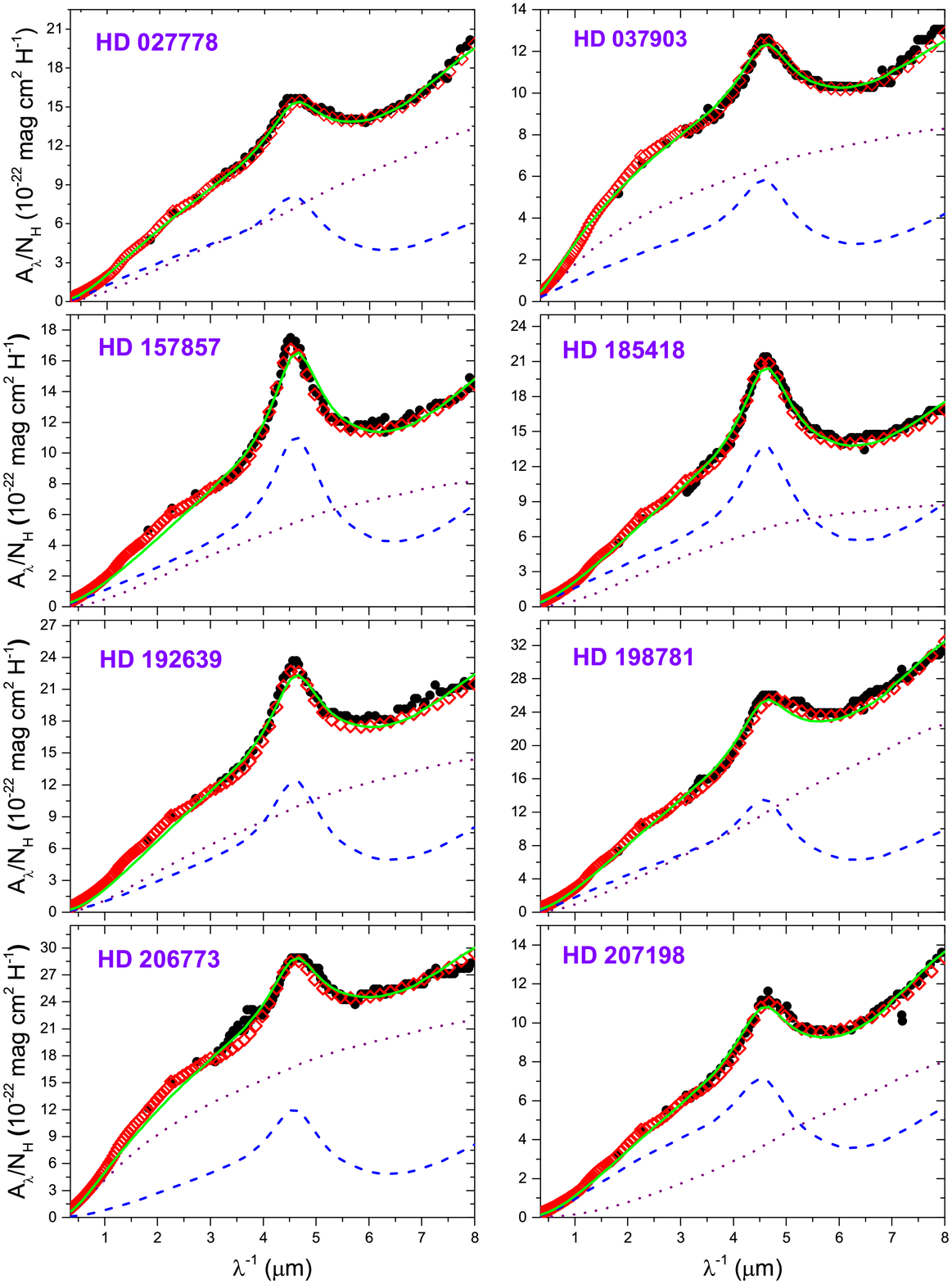}
\vspace{-2mm}
\caption{\footnotesize
         \label{fig:extmod1}
         Observed and model extinction curves of 
         HD\,027778, HD\,037903, HD\,157857,
         HD\,185418, HD\,192639, HD\,198781,
         HD\,206773, and HD\,207198.
         The observed extinction curves are represented
         by the FM90 parametrization 
         at $\lambda^{-1} > 3.3\mum^{-1}$
         and by the CCM parametrization 
         at $\lambda^{-1} < 3.3\mum^{-1}$
         (open red diamonds).
         The filled black circles plot
         the optical data at $U$, $B$, $V$ bands
         and the IUE data at $3.3 < \lambda^{-1} < 8.7\mum^{-1}$
         (taken from Gordon et al.\ 2009).
         The solid green line plots the model
         extinction curve which is a combination
         of silicate (dotted purple line)
         and graphite (dashed blue line).
         }
\end{figure*}

\begin{figure*}
\centering	
\vspace{-5mm}
\includegraphics[width=0.80\textwidth]{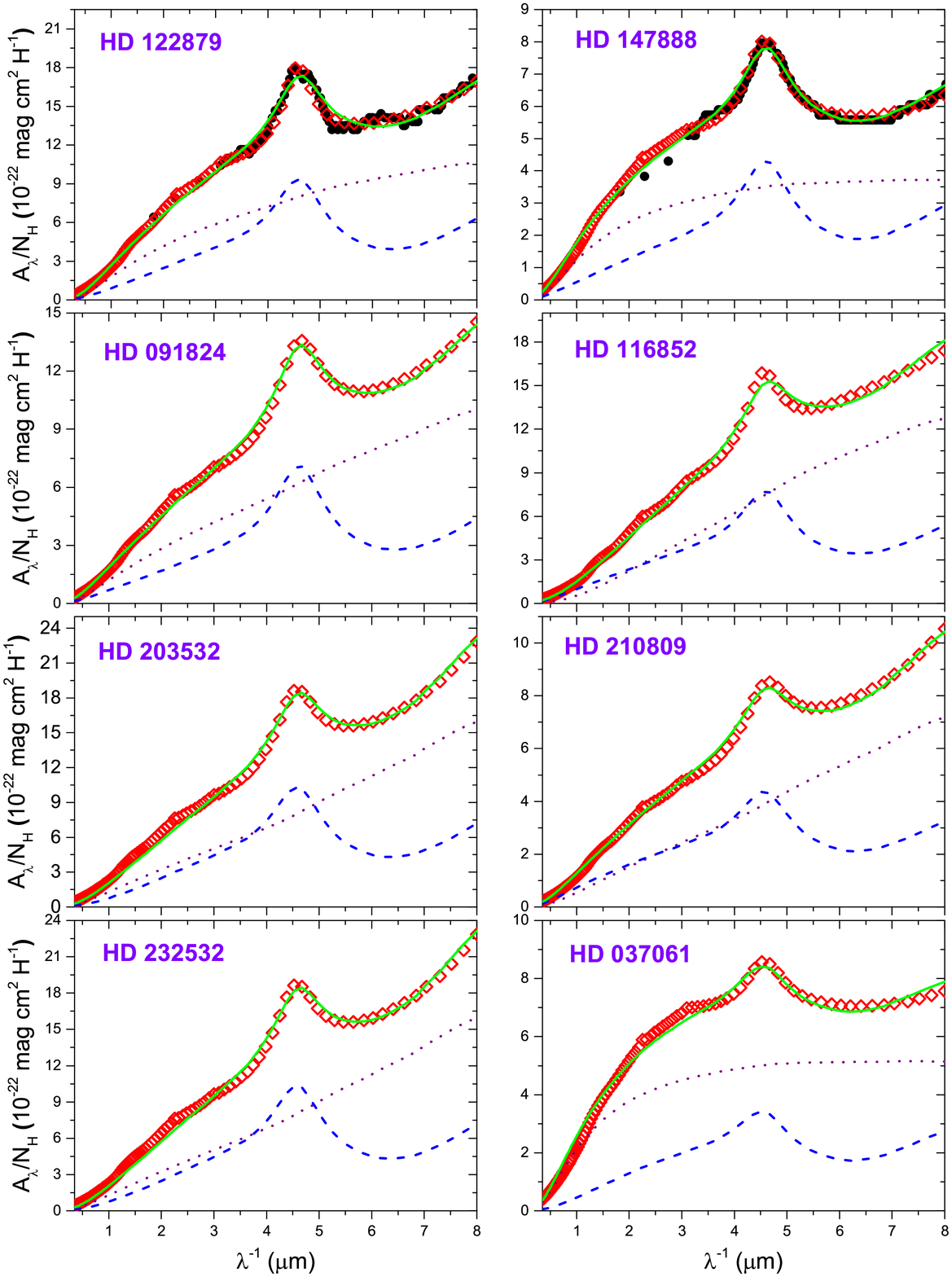}
\vspace{-2mm}
\caption{\footnotesize
         \label{fig:extmod2}
         Same as Figure~\ref{fig:extmod2}
         but for HD\,122879, HD\,147888,
         HD\,091824, HD\,116852, HD\,203532,
         HD\,210809, HD\,232532, and HD\,037061.
         For the latter six sources, the optical
         and IUE data are not shown.
         }
\end{figure*}

Let $\Aint \equiv \int_{0}^{\infty} A_\lambda/\NH\,d\lambda$ 
be the extinction integrated over all wavelengths.
The total volume of carbon dust per H nucleon
($\VC/\rmH$) can be derived from 
the Kramers-Kronig relation 
of Purcell (1969; see eq.\,\ref{eq:kk})
if we assume that the extinction is caused 
by silicate dust and carbon dust
\begin{equation}\label{eq:VC2H}
 \frac{\VC}{\rm H} = \left(\frac{1}{\FC}\right)
                     \left(\frac{\Aint}
                     {1.086\times 3 \pi^2} 
  -\Fsil\,\frac{\Vsil}{\rm H}\right) ~~~,
\end{equation}
where $\FC$ and $\Fsil$ are respectively 
the dimensionless factor $F$ seen in eq.\,\ref{eq:kk} 
for carbon dust and silicate dust. 
For moderately elongated grains,
$\Fsil\approx0.7$ for silicate materials
and $\FC\approx1.25$ for conducting, graphitic 
materials (see Figure~\ref{fig:dpol}).
For the selected 16 sources,
the extinction per H nucleon $A_\lambda/\NH$
is only known over a limited range of wavelengths.
%for $0.3 < \lambda^{-1} < 8.7\mum^{-1}$.
Therefore, we first use eq.\,\ref{eq:A2AV}
to extrapolate the UV extinction 
of Valencic et al.\ (2004)
to $\lambda=912\Angstrom$.\footnote{%
  Gordon et al.\ (2009) studied the extinction curves
  of 75 Galactic sightlines 
  obtained with 
  %the {\it Far Ultraviolet Spectroscopic Explorer} 
  FUSE at $905 < \lambda < 1187\Angstrom$ 
  and the {\it International Ultraviolet Explorer} (IUE) 
  at $1150 < \lambda < 3300\Angstrom$.  
  They found that the extrapolation
  of the UV extinction at $3.3 < \lambda^{-1} < 8.7\mum^{-1}$
  obtained by IUE
  is generally consistent with the far-UV extinction 
  at $8.4 < \lambda^{-1} < 11\mum^{-1}$
  obtained by FUSE.
  }
For $0.001 < \lambda^{-1} < 0.3\mum^{-1}$,
we adopt the model $A_\lambda/\NH$ values of
WD01 and Li \& Draine (2001b)
for the diffuse ISM.
This is justified since the near- and mid-IR
extinction at $\lambda > 0.9\mum$ does not
seem to vary much among different environments
(see Draine 2003a, Wang \& Jiang 2014, 
Wang et al.\ 2013, 2014).
In this way, we obtain 
$\Aintp = \int_{912\Angstrom}^{1000\mum} 
A_\lambda/\NH\,d\lambda$ for each source
(see Table~\ref{tab:VC2H}).
Since $\Aintp < \Aint$,
the volume of carbon dust per H nucleon ($\VC/\rmH$)
derived from eq.\,\ref{eq:VC2H} is a lower limit
if we replace $\Aint$ by $\Aintp$.
In the following, unless otherwise stated,
by $\Aint$ we really mean $\Aintp$.

What is the silicate dust volume per H nucleon
$\Vsil/\rmH$? As shown in \S\ref{sec:intro},
this depends on the assumed interstellar Fe,
Mg and Si abundances. In the literature, 
the proto-Sun abundances (Lodders 2003),
the solar abundances (Asplund et al.\ 2009),
the abundances of young F and G stars
(Sofia \& Meyer 2001) and/or
unevolved early B-type stars 
(Przybilla et al.\ 2008)
%, Nieva \& Przybilla 2012)
have been adopted to represent 
the interstellar standard abundances.
In this work we will only consider 
the proto-Sun abundances 
and the abundances of early B stars
since the solar abundances and
the abundances of young F and G stars 
are intermediate between that
of the proto-Sun and B stars
(see Footnote-\ref{ftnt:abund}).
The proto-Sun abundances of Fe, Mg and Si result in 
$\Vsil/\rmH \approx 3.17\times10^{-27}\cm^{3}\HH^{-1}$
while the B-star abundances lead to
$\Vsil/\rmH \approx 2.52\times10^{-27}\cm^{3}\HH^{-1}$.
Assuming the carbon dust to be graphite,\footnote{%
  As shown in Figure~\ref{fig:dpol}, 
  for moderately elongated or flattened spheroids
  the $\FC$ factors are essentially the same
  for graphite and amorphous carbon. 
  Therefore, the results derived from
  eq.\,\ref{eq:VC2H}
  also hold for amorphous carbon.
  }
we determine $\VC/\rmH$ from eq.\,\ref{eq:VC2H}
for each sightline.
The required carbon depletion is
\beq
\cdust\approx \left(\VC/\rmH\right)\rhoC/12\mH ~~,
\eeq
where $\rhoC$ is the mass density of the carbon dust 
($\rhoC\approx2.24\g\cm^{-3}$ for graphite).

In Table~\ref{tab:VC2H} we list the total carbon
volume $\VC/\rmH$ and carbon depletion $\cdust$
per H atom required to account for the observed
extinction for each sightline. 
For a given set of interstellar 
Fe, Mg and Si reference abundances,
the uncertainties on $\cdust$ mainly arise from
$\FC$, $\Fsil$ and $\siism$.
For moderately elongated or flattened spheroids,
the $\FC$ and $\Fsil$ values adopted here 
can vary at most $\simali$20\%.
For a given reference standard
(proto-Sun or B stars), the uncertainty on
$\siism$ is within $\simali$5\%.
Therefore, the uncertainties on
$\cdust$ are within $\simali$25\%.
We note that, 
with $A_V\approx2.16\pm0.04\magni$ (Gordon et al.\ 2009)
and $\NH\approx1.78\times10^{21}\cm^{-2}\HH$ 
(Cartledge et al.\ 2004),
the line of sight toward HD\,206773 is very dusty 
in spite it has a large gas-phase carbon abundance 
($\cgas\approx464\pm57\ppm$; Parvathi et al.\ 2012).
With $A_V/\NH\approx1.12\times10^{-21}\magni\cm^2\HH^{-1}$,
the HD\,206773 sightline is twice as dusty 
as the Galactic diffuse ISM of which
$A_V/\NH\approx5.3\times10^{-22}\magni\cm^2\HH^{-1}$.
Therefore, one requires an unusually large carbon
depletion ($\cdust\approx733\ppm$ if the proto-Sun
Fe, Mg, and Si abundances are adopted,
or $\cdust\approx774\ppm$ if the B-star
Fe, Mg, and Si abundances are adopted),
much larger than that of the other sightlines 
considered in this work.
In the following, unless otherwise stated,
we will not include HD\,206773 for correlation studies.

\begin{figure*}
\centering
\includegraphics[width=1.0\textwidth]{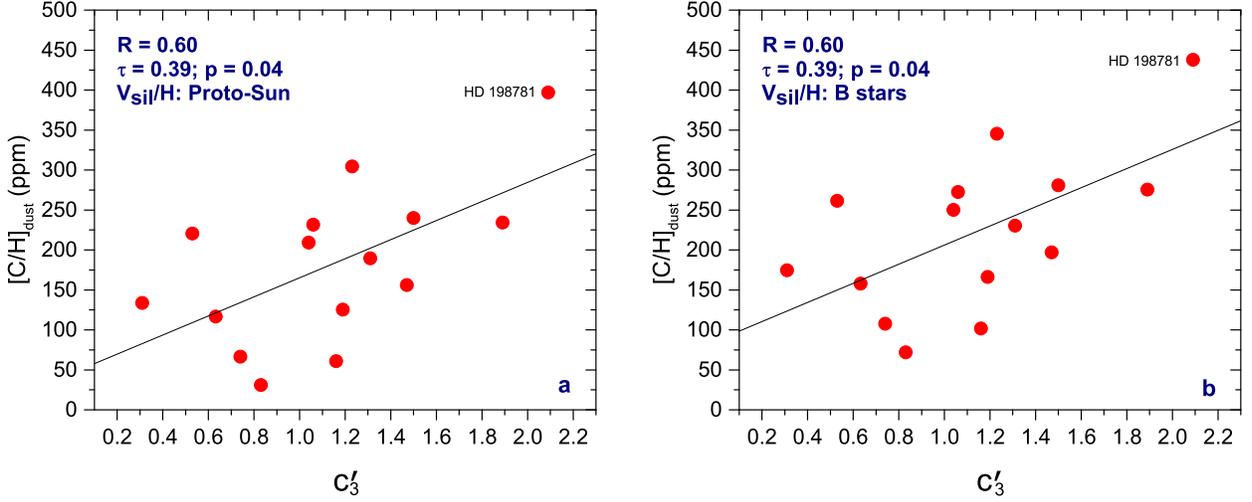}
\caption{\footnotesize
         \label{fig:c3_C2H}
         Correlation diagrams between the carbon depletion
         $\cdust$ with the strength of the 2175$\Angstrom$ 
         extinction bump ($c_3^{\prime}$). 
         The carbon depletion is derived from 
         the Kramers-Kronig relation 
         (see eqs.\,\ref{eq:kk},\ref{eq:VC2H}),
         with the interstellar Fe, Mg and Si
         abundances taken to be 
         that of proto-Sun (a) or B stars (b).
         The assumed interstellar Fe, Mg and Si
         reference abundances would not affect
         the correlation coefficients 
         but the intercept. 
         $R$ is the Pearson
	 correlation coefficient.
         Also labelled are the Kendall's $\tau$
	 coefficient and the significance level $p$.
         %... at 95$\%$ level of
	 %confidence ($\alpha$=0.05).
         }
\end{figure*}

\begin{figure*}
\centering
\includegraphics[width=1.0\textwidth]{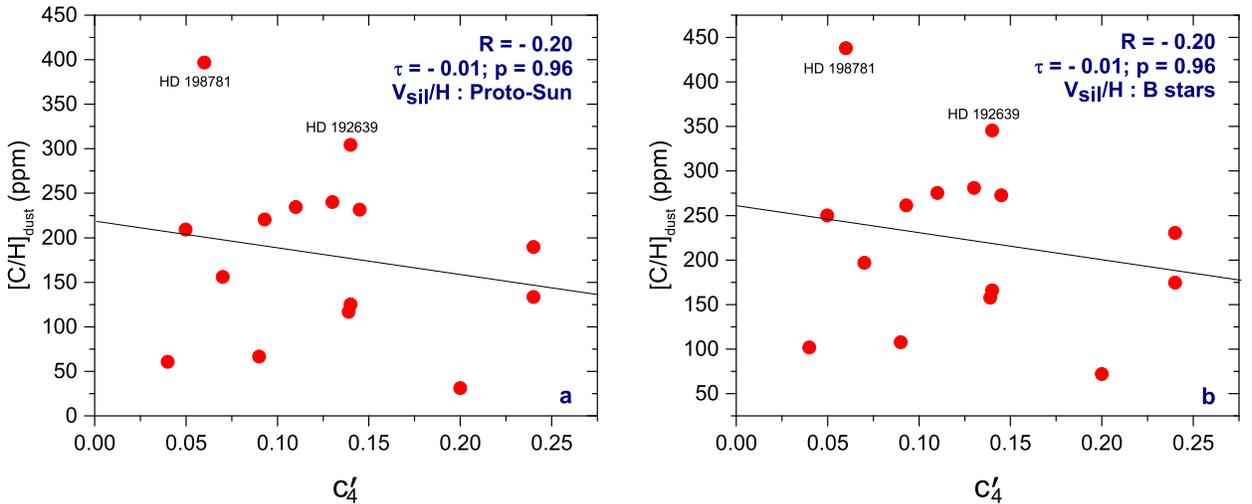}
\caption{\footnotesize
         \label{fig:c4_C2H}
         Correlation diagrams between the carbon depletion
         $\cdust$ with the strength of the far-UV nonlinear
         extinction rise ($c_4^{\prime}$). 
         The carbon depletion is derived from 
         the Kramers-Kronig relation 
         (see eqs.\,\ref{eq:kk},\ref{eq:VC2H}),
         with the interstellar Fe, Mg and Si
         abundances taken to be 
         that of the proto-Sun (a) or B stars (b).
         }
\end{figure*}

In Figure~\ref{fig:c3_C2H} we plot the carbon depletion
$\cdust$ as a function of $c_3^{\prime}$, the strength
of the 2175$\Angstrom$ extinction bump.
It can be seen from Figure~\ref{fig:c3_C2H} that 
$\cdust$ shows an increasing trend with $c_3^{\prime}$, 
with a Pearson linear correlation coefficient 
of $R\approx0.60$,
and a Kendall correlation coefficient 
of $\tau\approx0.39$ and a corresponding probability
$p\approx0.04$ of a chance correlation.\footnote{%
%a significance level
%of $p\approx0.04$.}\footnote{%
   We set the level of confidence to be 95\%
   (i.e., Kendall's $\alpha=0.05$).
   }
This is independent of the
assumed interstellar Fe, Mg and Si reference abundances
which would only affect the intercept.
In Figure~\ref{fig:c4_C2H} we plot the carbon depletion
$\cdust$ as a function of $c_4^{\prime}$, the strength
of the nonlinear far-UV extinction rise.
Figure~\ref{fig:c4_C2H} shows no detectable correlation
between $\cdust$ and $c_4^{\prime}$
for which the Pearson correlation coefficient 
is $R\approx-0.20$
and the Kendall $\tau$ is $\approx-0.01$ 
and $p\approx0.96$.
Again, these correlation coefficients
are independent of the assumed interstellar 
Fe, Mg and Si reference abundances.

\section{Carbon Depletion Inferred from 
         the Extinction Modeling\label{sec:extmod}}
The carbon depletion $\cdust$ derived from
the Kramers-Kronig relation is independent of
any dust models except we just need to assume
that the observed extinction is caused by
silicate dust and carbon dust.\footnote{%
  The correlation coefficients derived in
  \S\ref{sec:kk} and shown in Figure~\ref{fig:c3_C2H}
  remain the same for both graphite and 
  amorphous carbon.
  The required carbon depletion $\cdust$ 
  would decrease by a factor of 
  $\rhogra/\rho_{\rm ac}$,
  where $\rho_{\rm ac}\approx1.8\g\cm^{-3}$ 
  is the mass density of amorphous carbon.
  }

We now would like to derive the carbon depletion
from fitting the observed extinction curve 
for each sightline.
We consider the silicate-graphite interstellar
grain model which consists of two separate dust
components: amorphous silicate and graphite
(Mathis et al.\ 1977, Draine \& Lee 1984).
We adopt an exponentially-cutoff power-law size
distribution for both components:
$dn_i/da = \nH B_i a^{-\alpha_i} \exp\left(-a/a_{c,i}\right)$
for the size range of
$50\Angstrom < a < 2.5\um$,
where $a$ is the spherical radius of the dust,
$\nH$ is the number density of H nuclei,
$dn_i$ is the number density of dust of type $i$
with radii in the interval [$a$, $a$\,$+$\,$da$],
$\alpha_i$ and $a_{c,i}$ are respectively
the power index and exponential cutoff size
for dust of type $i$, and
$B_i$ is the constant related to
the total amount of dust of type $i$.
The total extinction per H column
at wavelength $\lambda$ is given by
\begin{equation}\label{eq:Amod}
A_\lambda/\NH = 1.086
            \sum_i \int da \frac{1}{\nH} 
            \frac{dn_i}{da}
            C_{{\rm ext},i}(a,\lambda),
\end{equation}
where the summation is over the two grain types
(i.e., silicate and graphite),
$\NH\equiv\int\nH\,dl$ is the H column density
which is the H number density
integrated over the line of sight $l$,
and $C_{{\rm ext},i}(a,\lambda)$
is the extinction cross section of
grain type $i$ of size $a$
at wavelength $\lambda$
which can be calculated 
from Mie theory (Bohren \& Huffman 1983)
using the dielectric functions of
``astronomical'' silicate and graphite 
of Draine \& Lee (1984).

In fitting the extinction curve,
for a given sightline, 
we have six parameters:
the size distribution power indices 
$\alphaS$ and $\alphaC$ 
for silicate and graphite, respectively;
the exponential cutoff sizes 
$\acS$ and $\acC$, respectively;
and $\Bsil$ and $\Bgra$.
We derive the silicon and carbon depletions from
\begin{equation}\label{eq:Si2H}
\sidust = \left(\nH\Bsil/172\mH\right) 
        \int da \left(4\pi/3\right) a^3\,\rhosil
               a^{-\alphaS}
               \exp\left(-a/\acS\right) ~~,
\end{equation}
\begin{equation}\label{eq:C2H}
\cdust = \left(\nH\Bgra/12\mH\right) 
        \int da \left(4\pi/3\right) a^3\,\rhogra
               a^{-\alphaC}
               \exp\left(-a/\acC\right) ~~,
\end{equation}
where we assume a stoichiometric composition of
MgFeSiO$_4$ for amorphous silicate.

For a given sightline,
we seek the best fit to the extinction 
between $0.3\mum^{-1}$ and $8\mum^{-1}$
by varying the size distribution power indices 
$\alphaS$ and $\alphaC$, and
the upper cutoff size parameters 
$\acS$ and $\acC$.
Following WD01,
we evaluate the extinction at 
100 wavelengths $\lambda_i$, 
equally spaced in $\ln\lambda$.
We use the Levenberg-Marquardt method
(Press et al.\ 1992) to minimize $\chi^2$ 
which gives the error in the extinction fit:
\begin{equation}
\chi^2 = \sum_i \frac{\left( \ln A_{\rm obs} 
       - \ln A_{\rm mod} \right)^2} {\sigma_i^2}~~~,
\end{equation}
where $A_{\rm obs}(\lambda_i)$ 
is the observed extinction at wavelength $\lambda_i$,
$A_{\rm mod}(\lambda_i)$ 
is the extinction computed
for the model at wavelength $\lambda_i$
(see eq.\,\ref{eq:Amod}), 
and the $\sigma_i$ are weights.  
Following WD01,
we take the weights $\sigma_i^{-1} = 1$ 
for $1.1 < \lambda^{-1} < 8\mum^{-1}$ 
and $\sigma_i^{-1} = 1/3$ for 
$\lambda^{-1} < 1.1\mum^{-1}$.

\begin{figure*}
\centering
\includegraphics[width=1.0\textwidth]{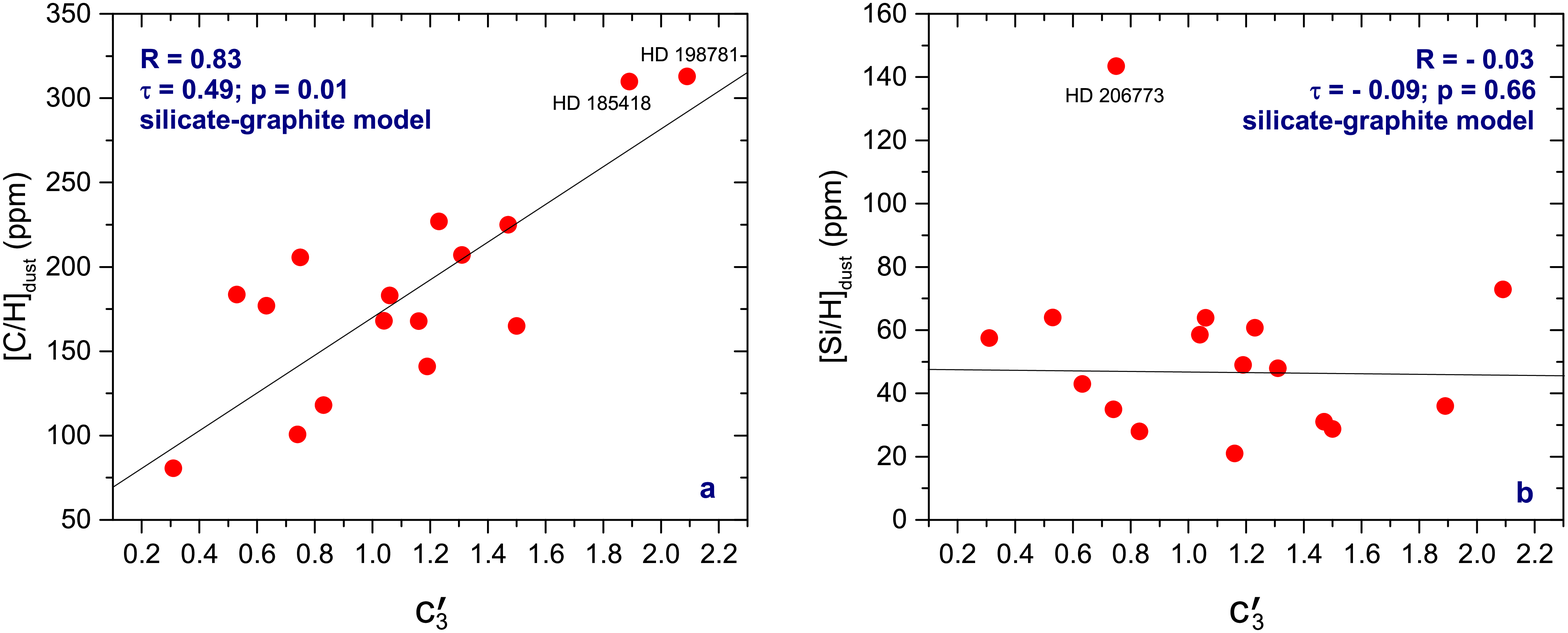}
\caption{\footnotesize
         \label{fig:c3_vs_C_Si_mod}
         Correlation diagrams between the strength of 
         the 2175$\Angstrom$ extinction bump ($c_3^{\prime}$) 
         with the carbon depletion $\cdust$ (a) 
         and silicon depletion $\sidust$ (b) 
         derived from fitting the extinction 
         of each sightline with a mixture of
         silicate dust and graphite dust
         (see \S\ref{sec:extmod}).
         }
\end{figure*}

\begin{figure*}
\centering
\includegraphics[width=1.0\textwidth]{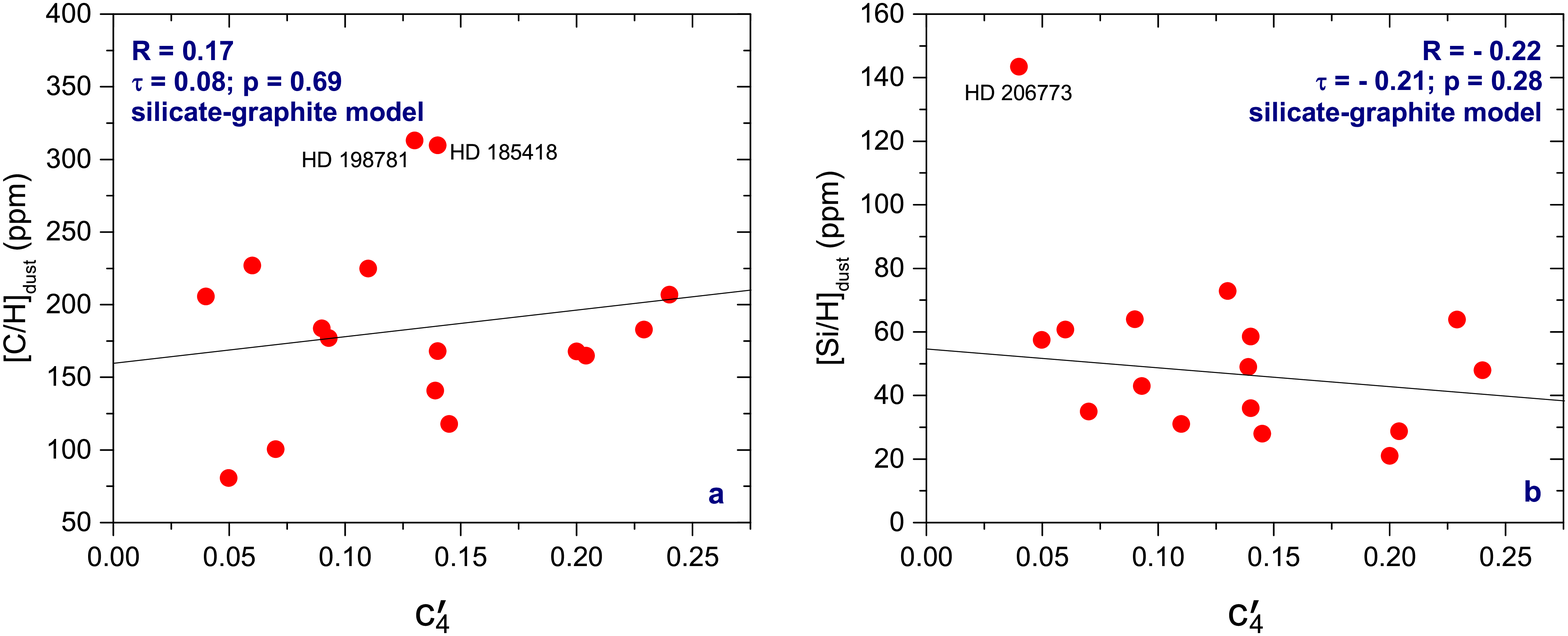}
\caption{\footnotesize
         \label{fig:c4_vs_C_Si_mod}
         Correlation diagrams between the strength of 
         the nonlinear far-UV extinction rise ($c_4^{\prime}$) 
         with the carbon depletion $\cdust$ (a) 
         and silicon depletion $\sidust$ (b) 
         derived from fitting the extinction 
         of each sightline with a mixture of
         silicate dust and graphite dust
         (see \S\ref{sec:extmod}).
         }
\end{figure*}

In Figures~\ref{fig:extmod1},\,\ref{fig:extmod2}
we show the model fits. It can be seen from these 
figures that a simple mixture of silicate and graphite
closely reproduces the observed UV/optical/near-IR 
extinction of all 16 sightlines.
The model parameters are tabulated in Table~\ref{tab:modpara}.
In Figure~\ref{fig:c3_vs_C_Si_mod} we examine 
the correlations between the strength of 
the 2175$\Angstrom$ extinction bump ($c_3^{\prime}$) 
with the carbon ($\cdust$) 
and silicon ($\sidust$) depletions
derived from fitting the observed extinction
(see eqs.\,\ref{eq:Si2H},\,\ref{eq:C2H}).
With a Pearson correlation coefficient 
of $R\approx0.83$
and a Kendall $\tau\approx0.49$ 
and $p\approx0.01$,
we see that the carbon depletion correlates 
with the 2175$\Angstrom$ bump, 
while the silicon depletion does not 
correlate with the 2175$\Angstrom$ bump. 
We have also explored the relations
between the strength of the nonlinear far-UV
extinction rise ($c_4^{\prime}$) 
and the carbon and silicon depletions.
As shown in Figure~\ref{fig:c4_vs_C_Si_mod},
we do not see any correlation.

\section{Discussion\label{sec:discussion}}
Interstellar dust plays an important role 
in various physical and chemical processes
that take place in the ISM. 
Dust extinction remains the most direct way to 
investigate the properties of interstellar dust
since Trumpler (1930) first established the existence 
of solid particles in interstellar space 
through the discovery of color excess.  
However, practically most of the extinction modeling 
efforts have been so far directed to the Galactic average 
extinction curve which is obtained by averaging over many 
clouds of different gas and dust properties.
An investigation of dust extinction toward individual 
clouds is of special importance.

\begin{figure*}
\centering
\includegraphics[width=1.0\textwidth]{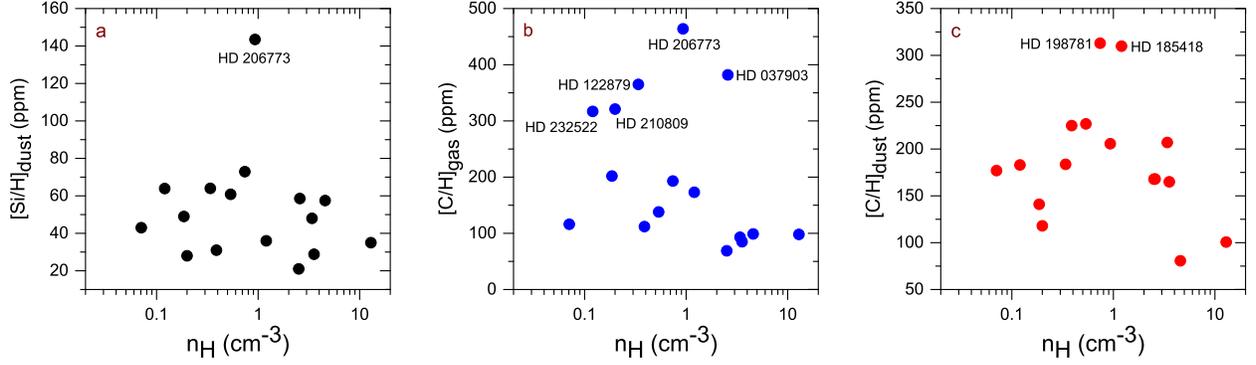}
\caption{\footnotesize
         \label{fig:nH_Si_C}
         Silicon ($\sidust$) and carbon ($\cdust$) depletions 
         as well as gas-phase carbon abundance ($\cgas$)
         plotted as a function of the average hydrogen 
         density $\langle\nH\rangle$.
         The $\sidust$ and $\cdust$ depletions 
         are derived from fitting the extinction 
         of each sightline with a mixture of
         silicate dust and graphite dust
         (see \S\ref{sec:extmod}).
         }
\end{figure*}

In this work we explore the extinction and dust depletion
of 16 individual sightlines of which $R_V$ ranges 
from $\simali$2.4 to $\simali$4.3 and the average
hydrogen density $\langle\nH\rangle$ ranges over
nearly three orders of magnitudes 
(see Figure~\ref{fig:nH_Si_C}).
The hydrogen column density 
and the gas-phase carbon abundance 
are known for these sightlines,
with the latter determined from
the strong 1334$\Angstrom$ CII absorption line 
(Parvathi et al.\ 2012, Sofia et al.\ 2011).
Unlike Cardelli et al.\ (1996) who derived 
the interstellar gas-phase carbon abundance 
to be $\cgas\approx\left(140\pm20\right)\ppm$
and found that $\cgas$ shows no dependence 
on the physical condition of the gas, 
for the sightlines studied by
Parvathi et al.\ (2012) and Sofia et al.\ (2011),
the gas-phase $\cgas$ appears to show a decreasing
trend for increasing values of $\langle\nH\rangle$
(see Figure~\ref{fig:nH_Si_C}, also see Figure~3 
in Parvathi et al.\ 2012).\footnote{%
   Cardelli et al.\ (1996) determined $\cgas$
   from the weak intersystem line of CII] 
   at 2325$\Angstrom$ obtained with 
   the HST {\it Goddard High Resolution Spectrograph}
   (GHRS) for six sightlines toward stars within 600\,pc 
   of the Sun. These sightlines include examples 
   exhibiting a wide range of extinction variation.
   }

The lack of any statistically significant variation 
in the observed gas-phase $\cgas$ 
among the six sightlines studied 
by Cardelli et al.\ (1996)
is in strong contrast to 
the appreciable variation of $\cgas$ 
with $\langle\nH\rangle$ seen in 
the sightlines studied in this work.
As shown in Figure~\ref{fig:nH_Si_C},
the solid-phase silicon ($\sidust$) 
and carbon ($\cdust$) abundances 
derived from fitting the observed extinction
also exhibit substantial variations with 
$\langle\nH\rangle$.
The solid-phase carbon abundance ($\cdust$) 
determined from the wavelength-integrated 
extinction ($\Aint$) based on the Kramers-Kronig 
relation varies by a factor of $>$10 among 
these 16 sightlines.
Therefore, the total abundance of carbon,
the sum of the components residing in the gas
and in the dust, varies significantly
among these sightlines. This raises a question:
{\it is there indeed a reference standard for 
the interstellar carbon abundance?}

As shown in Table~\ref{tab:VC2H}, 
the analysis of the wavelength-integrated extinction
based on the Kramers-Kronig relation 
shows that five of the 16 sources require 
in total a carbon abundance 
(i.e., $\cdust$\,$+$\,$\cgas$)
exceeding $\simali$500$\ppm$,
much higher than that of 
the young F and G stars 
($\cstar\approx355\pm82\ppm$, Sofia \& Meyer 2001),
the proto-Sun 
($\csun\approx288\pm27\ppm$, Lodders 2003),
the Sun 
($\csun\approx269\pm31\ppm$, Asplund et al.\ 2009),
and the early B stars
($\cstar\approx214\pm20\ppm$, Przybilla et al.\ 2008),
%$\cstar\approx209\pm15\ppm$, Nieva \& Przybilla 2012),
which are often taken to be the interstellar carbon
reference abundance.  
This is also true for the carbon depletion
derived from fitting the observed extinction
with the silicate-graphite model.
%
%If there exists a reference standard,
%then the carbon reference abundance needs 
%to exceed $\simali$500$\ppm$.
%In this case, although the total amount
%of carbon can account for the observed
%extinction and $\cgas$ for some sightlines,
%but it cannot be all accounted for other sightlines. 
%
Mathis (1996) argued that a dust model may
require fewer C, Si, Mg and Fe atoms to account
for the observed extinction if the dust has a fluffy, 
porous structure since fluffy grains are more effective 
in absorbing and scattering the UV/optical starlight than 
compact grains (on a per unit mass basis). 
However, as demonstrated in Li (2005),
fluffy dust is not able to appreciably 
reduce the consumption of dust-forming atoms.
As can be seen in eq.\,\ref{eq:kk},
although the dust volume $V_{\rm dust}$ does
increase when the dust has voids and becomes porous,
the $F$-factor decreases and therefore, for a given
amount of dust material, the wavelength-integrated 
extinction $\int A_\lambda/\NH\,d\lambda$ does not
change much (see Figures~1,2 in Li 2005). 
Alternatively, one may argue that the gas-phase $\cgas$ 
abundances derived from the strong CII absorption line 
at $\lambda\approx1334\Angstrom$ may be overestimated 
since accurate measures of the absorption 
in the wings of this line is very difficult. 
However, the weak CII line at $\lambda\approx2325\Angstrom$  
gives even higher $\cgas$ abundances
(Sofia et al.\ 2004, Parvathi et al.\ 2010).
Therefore, it seems real that the total C/H
abundance of some of our 16 sigtlines is well above 
that of the popular interstellar reference standards.

In fitting the observed extinction 
(see \S\ref{sec:extmod} and Table~\ref{tab:modpara}), 
seven out of the 16 sources require $\sidust$ to exceed
the proto-Sun silicon abundance by $>$50\%,
three of them are within $\simali$20\% of 
the proto-Sun silicon abundance,
and six of them require less silicon 
than that of the proto-Sun.
%If it is true that silicon is essentially
%fully depleted in dust in the ISM,
%the variation in $\sidust$ seen in these 16
%sightlines indicates that the total silicon
%abundance may vary from one region to another,
%i.e., there are no reference standards for
%the interstellar silicon abundance.
%%In this case, the carbon depletion $\cdust$
%%derived from the Kramers-Kronig relation
%%(see \S\ref{sec:kk} and Table~\ref{tab:VC2H})
%%should be treated with caution since an interstellar
%%standard reference of the proto-Sun or Sun
%%has been assumed for Si, Mg and Fe. 
%%%
%%This probably also explains why Parvathi et al.\ (2012)
%%could not find a positive correlation between the carbon
%%depletion $\cdust$ and the strength of the 2175$\Angstrom$
%%extinction bump ($c_3^{\prime}$): they assumed a reference
%%standard of $\cism = 464\ppm$ for all sightlines.
%%cosmic carbon abundance.
%%
%%Nevertheless, 
The variations 
in the silicon and carbon depletions 
toward different sightlines %does 
imply various physical processing 
(e.g., growth, coagulation, destruction, 
and photo-processing) of interstellar dust
and variations in the composition 
within the Milky Way (Parvathi et al.\ 2012).

\begin{figure*}
\centering
\includegraphics[width=1.0\textwidth]{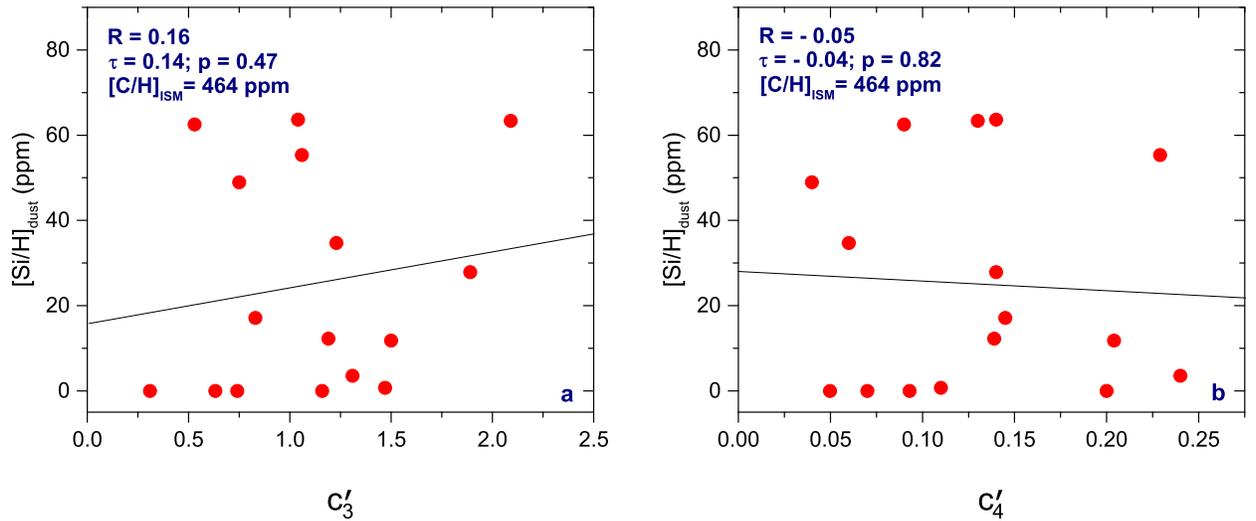}
\caption{\footnotesize
         \label{fig:c3_Si_mod}
         Correlation diagrams between
         the Si/H depletion $\sidust$ 
         with the strength of the extinction bump
         ($c_3^{\prime}$; a) 
         and with the far-UV extinction rise 
	 ($c_4^{\prime}$; b). 
         The Si/H depletion
         is derived from the Kramers-Kronig relation, 
	 with the total C/H abundance fixed
         at $\cism=464\ppm$.
         }
\end{figure*}

Parvathi et al.\ (2012) found no correlation 
between the carbon depletion $\cdust$ and 
the strength of the 2175$\Angstrom$
extinction bump ($c_3^{\prime}$), 
while in this work we find $\cdust$ 
correlates with $c_3^{\prime}$
(see \S\ref{sec:kk} and \S\ref{sec:extmod}). 
The major difference between our approaches
and Parvathi et al.\ (2012) is that they
assumed a reference standard of $\cism = 464\ppm$ 
for all sightlines, 
while the approaches taken by us 
which make use of the Kramers-Kronig relation 
(see \S\ref{sec:kk})
or model the observed extinction (see \S\ref{sec:extmod})
result in considerable
variation in the total C/H abundance. 
We note that for the Kramers-Kronig-relation-based approach 
we have assumed a reference standard for Si, Mg and Fe 
(see \S\ref{sec:kk}). One may argue that 
the correlation between $\cdust$ and $c_3^{\prime}$
derived from this approach may not be real:
with $\sidust$, $\mgdust$ and $\fedust$ 
fixed at a reference standard,
one might expect the required C/H depletion
to increase with the extinction.
However, this is not true. 
With $\sidust$, $\mgdust$ and $\fedust$ fixed,
we do expect $\cdust$ to increase with $A_\lambda/\NH$,
but we do not necessarily expect $\cdust$ to
increase with $c_3^{\prime}$ since the latter
is a quantity normalized by $A_V$ and therefore
the common correlation of $\cdust$ 
and $c_3^{\prime}$ with $A_V$ is cancelled.
This can be tested by fixing $\cism$ 
and determining the Si/H depletion from
the Kramers-Kronig relation and then examining 
the correlation between $c_3^{\prime}$ and $\sidust$.
Assuming the same total C/H abundance of $\cism$ for
all 16 sightlines, the carbon dust volume per H nucleon
for a given sightline can be derived from its 
gas-phase C/H abundance:
\begin{equation}
 \frac{\VC}{\rm H} = 
\left\{\cism-\cgas\right\}\times 12\,\mH/\rho_{\rm C} ~~.
\end{equation}
We use the Kramers-Kronig relation to
determine the silicate volume per H nucleon
from the wavelength-integrated extinction:
\begin{equation}
 \frac{\Vsil}{\rm H} = \left(\frac{1}{\Fsil}\right)
                     \left(\frac{\Aint}
                     {1.086\times 3 \pi^2} 
  -\FC\,\frac{\VC}{\rm H}\right) ~~~.
\end{equation}
Then we derive the required Si/H depletion from
\beq
\sidust = \left(\Vsil/\rmH\right)
          \rho_{\rm sil}/\mu_{\rm sil}\mH ~~,
\eeq
where $\mu_{\rm sil}=172$ is the molecular weight 
for MgFeSiO$_4$. 
Following Parvathi et al.\ (2012),
we adopt $\cism=464\ppm$.
In Figure~\ref{fig:c3_Si_mod} we plot
$\sidust$ against $c_3^{\prime}$ 
and $c_4^{\prime}$ and see no correlation.
This justifies the correlation between
$\cdust$ and $c_3^{\prime}$ derived
in the Kramers-Kronig-relation-based approach
presented in \S\ref{sec:kk} which assumes
fixed $\sidust$, $\mgdust$ and $\fedust$ depletions.

We note that although the carbon abundance $\cism$ is 
underabundant in accounting for the observed extinction,
no matter what reference standard (proto-Sun, Sun,
early B stars, or young F and G stars) is adopted,
the oxygen abundance remains largely unaccounted for
(see Whittet 2010). Recently, Poteet et al.\ (2015)
argued that the missing oxygen may reside in large,
micrometer-sized H$_2$O ice grains which are nearly 
opaque to IR radiation. 
Possible evidence for 
the presence of $\mu$m-sized 
interstellar grains have been found 
by Wang et al.\ (2014), 
who modeled the mid-IR extinction toward 
a variety of interstellar environments 
using a population of $\mu$m-sized grains.
Furthermore, the recent analysis of three 
possibly interstellar grains returned by 
the {\it Stardust} spacecraft also provides 
tentative evidence for the presence of
$\mu$m-sized grains in the ISM 
(Sterken et al.\ 2014, Westphal et al.\ 2014).
These $\mu$m-sized grains, although gray in
the UV/optical/near-IR, will contribute 
to the wavelength-integrated extinction $\Aint$ 
and therefore reduce the consumption
of silicon and carbon 
(i.e., decreasing $\cdust$ and $\sidust$).

The correlation between the 2175$\Angstrom$ bump strength 
and the carbon depletion $\cdust$ 
inferred both from the
%model-independent 
Kramers-Kronig relation 
and from fitting the observed extinction with a specific 
dust model supports the hypothesis of 
carbonaceous grains 
(e.g., graphite or PAHs)
as the possible carriers of 
the 2175$\Angstrom$ bump.\footnote{%
   Amorphous carbon (Bussoletti et al.\ 1987), 
   graphitized %(dehydrogenated) 
   hydrogenated amorphous carbon 
   (HAC; Mennella et al.\ 1996), 
   and nano-sized HAC (Schnaiter et al.\ 1998)
   have also been proposed
   as candidate materials.
   However, they are not always successful in
   reproducing the observed invariant peak position
   and variable strength and width of 
   the 2175$\Angstrom$ bump.
   }
The lack of correlation between the silicon depletion
and the 2175$\Angstrom$ bump argues against small silicates
or (Mg, Si) oxides as its carrier (see Duley 1985).  

\begin{figure*}
\centering
\includegraphics[width=1.0\textwidth]{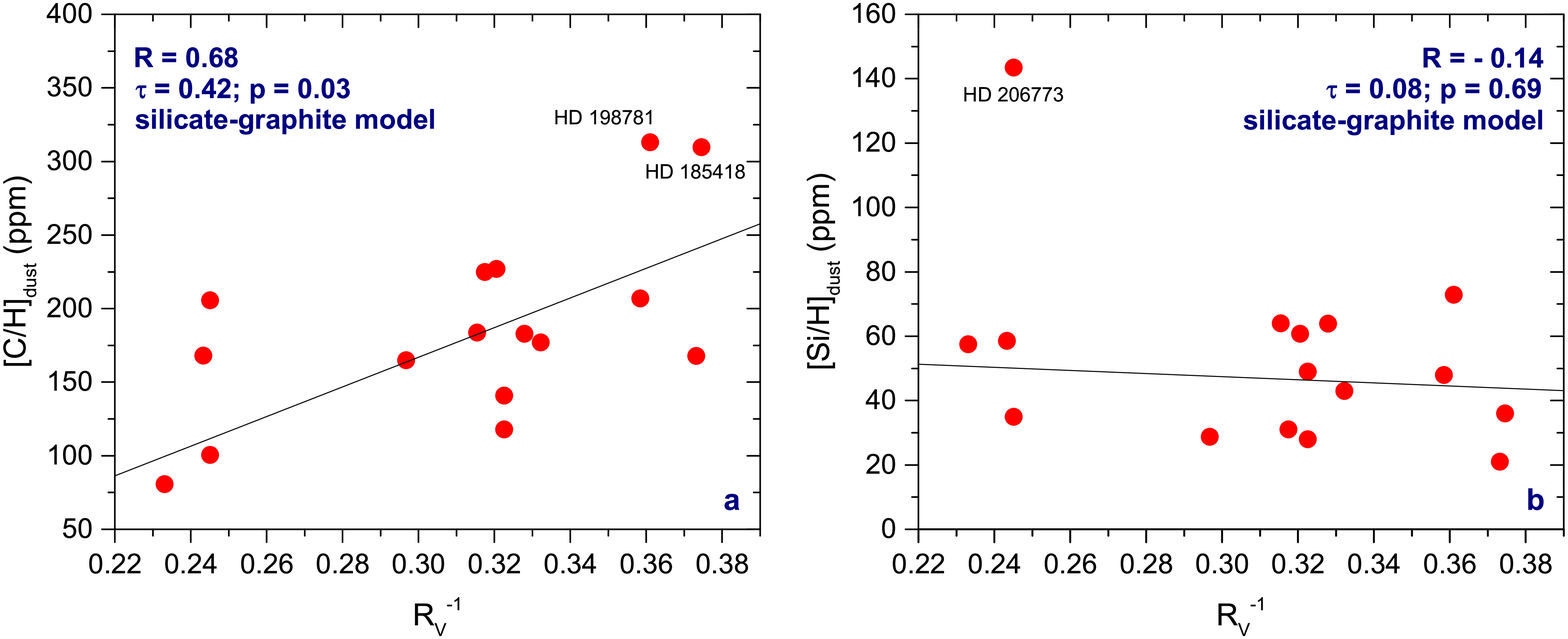}
\caption{\footnotesize
         \label{fig:RV_C_Si}
         Correlation diagrams between $R_V^{-1}$
         and the carbon depletion $\cdust$ (a) 
         and silicon depletion $\sidust$ (b) 
         derived from fitting the extinction 
         of each sightline with a mixture of
         silicate dust and graphite dust
         (see \S\ref{sec:extmod}).
          }
\end{figure*}

\begin{figure*}
\centering
\includegraphics[width=1.0\textwidth]{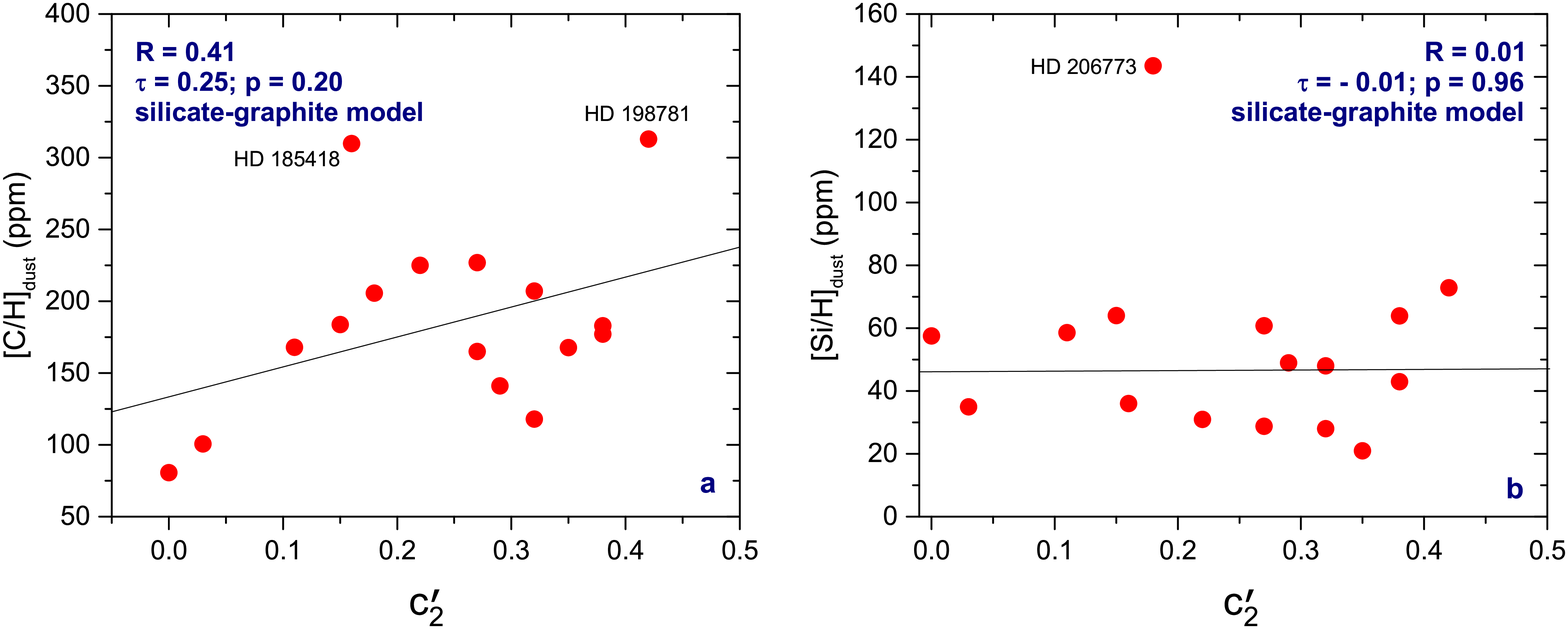}
\caption{\footnotesize
         \label{fig:c2_vs_C_Si_mod}
         Correlation diagrams between the linear
	 background ($c_2^{\prime}$) 
         with the carbon depletion $\cdust$ (a) 
         and silicon depletion $\sidust$ (b) 
         derived from fitting the extinction 
         of each sightline with a mixture of
         silicate dust and graphite dust.
         }
\end{figure*}

Finally, we examine the relations between the carbon and 
silicon depletions with $R_V^{-1}$. 
As shown in Figure~\ref{fig:RV_C_Si}, the carbon depletion
appears to moderately correlate with $R_V^{-1}$. 
$R_V$ is roughly indicative of the dust size: 
the smaller $R_V$ is,
the steeper is the far-UV extinction 
and the richer in small dust is 
the dust size distribution. 
This suggests that the far-UV extinction 
is more likely produced by small carbon dust
than by small silicate dust,
probably arising from 
the blue wing of 
the $\sigma$--$\sigma^{\ast}$
electronic absorption profile 
of graphitic materials.
We note that $R_V$ provides a better 
characterization of the far-UV extinction
than $c_4^{\prime}$ as the latter only
accounts for the far-UV {\it nonlinear} 
extinction rise while the {\it linear} background 
$c_1^{\prime} + c_2^{\prime} x$
surely also contributes to the far-UV extinction.
In Figure~\ref{fig:c2_vs_C_Si_mod}
we plot the C/H and Si/H depletions
against $c_2^{\prime}$.
As expected, we do not see any correlation
because, similar to $c_4^{\prime}$,  
$c_2^{\prime}$ alone does not
characterize the far-UV extinction.

\section{Conclusion\label{sec:conclusion}}
We have studied the extinction and dust depletion
in 16 Galactic sightlines 
in order to probe the role of carbon 
in the UV extinction, 
with special attention paid to 
the 2175$\Angstrom$ extinction bump.
%and the far-UV extinction rise.
These sightlines have their UV/optical/near-IR
extinction, gas-phase carbon abundances,
and hydrogen column densities well determined,
therefore allowing us to quantitatively 
explore the relations between the silicon
and carbon depletions and the UV extinction.
Our principal results are as follows:
\begin{enumerate}
\item We have derived the carbon depletion $\cdust$
      from the Kramers-Kronig relation which
      relates the wavelength-integrated extinction
      to the total dust volume. 
      We find that $\cdust$ intends to correlate
      with the strength of the 2175$\Angstrom$ 
      extinction bump. %This is independent of
      %any dust models.
      This is less model-dependent
      since it does not require the knowledge 
      of the detailed optical properties and
      size distribution of the dust. 
\item We have derived the carbon $\cdust$ and 
      silicon $\sidust$ depletions 
      from fitting the observed
      UV/optical/near-IR extinction with 
      a mixture of silicate dust and 
      carbonaceous dust.
      We find that $\cdust$ correlates
      with the 2175$\Angstrom$ extinction bump
      and $\sidust$ does not correlate with 
      the 2175$\Angstrom$ bump.
      This supports graphite or PAH molecules 
      as the possible carrier of 
      the 2175$\Angstrom$ bump,
      while argues against the hypothesis
      of silicate or (Mg, Si) oxides  
      as its carrier.
\item A moderate correlation between $\cdust$ and $1/R_V$
      is derived from the extinction modeling.
      This suggests that the far-UV extinction 
      is more likely produced by small carbon dust 
      than by small silicate dust.
\end{enumerate}

\acknowledgments
We thank B.W.~Jiang, Q.~Li, B.~Mills, J.Y.~Seok, and Y.X.~Xie 
for helpful discussions.
We are particularly grateful to 
S. Federman, S.~Wang,
and the anonymous referee
for their suggestions which 
significantly improved the quality of this paper.
We are supported in part by
NSF AST-1109039, 
NNX13AE63G, and NSFC\,11173019, 11273022
and the University of Missouri Research Board.

% -------------------------------------------------------------------------

%%%%%%%%%%%%%%%%% Figure 1: F-factor %%%%%%%%%%%%%%%%%
%%%%%%%%%%%%%%%%% Figure 1: F-factor %%%%%%%%%%%%%%%%%

%%%%%%%%%%%%% Figure 2: A_lambda/N_H (1) %%%%%%%%%%%%%
%%%%%%%%%%%%% Figure 2: A_lambda/N_H (1) %%%%%%%%%%%%%

%%%%%%%%%%%%% Figure 3: A_lambda/N_H (2) %%%%%%%%%%%%%
%%%%%%%%%%%%% Figure 3: A_lambda/N_H (2) %%%%%%%%%%%%%

%%%%%%%%%%%% Figure 4: C/H vs. c_3 (2175A) %%%%%%%%%%%%
%%%%%%%%%%%% Figure 4: C/H vs. c_3 (2175A) %%%%%%%%%%%%

%%%%%%%%%%%% Figure 5: C/H vs. c_4 (far-UV) %%%%%%%%%%%%
%%%%%%%%%%%% Figure 5: C/H vs. c_4 (far-UV) %%%%%%%%%%%%

%%%%%%  Figure 6: Model C/H, Si/H vs. c_3 (2175A) %%%%%%
%%%%%%  Figure 6: Model C/H, Si/H vs. c_3 (2175A) %%%%%%

%%%%%%  Figure 7: Model C/H, Si/H vs. c_4 (far-UV) %%%%%%
%%%%%%  Figure 7: Model C/H, Si/H vs. c_4 (far-UV) %%%%%%

%%%%%%  Figure 8: n_H vs C/H, Si/H %%%%%%
%%%%%%  Figure 8: n_H vs C/H, Si/H %%%%%%

%%% Figure 9: c_3, c_4 vs Si/H with [C/H]_ISM = 464ppm fixed %%%
%%% Figure 9: c_3, c_4 vs Si/H with [C/H]_ISM = 464ppm fixed %%%

%%%% Figure 10 %%%%%
%%%% Figure 10 %%%%%

%%%%%%  Figure 11: c_2 vs C/H, Si/H %%%%%%
%%%%%%  Figure 11: c_2 vs C/H, Si/H %%%%%%

%%%%%%%%%%%%%%   Table 1   %%%%%%%%%%%%%%
\thispagestyle{empty}
\setlength{\voffset}{45mm}
\begin{deluxetable}{lccccccccccccccccr}
\rotate 
\tablecolumns{16}
\tabletypesize{\tiny}
\tablewidth{0truein}
\center
\tablecaption{\footnotesize
              \label{tab:VC2H}
              Extinction Parameters 
              and the Carbon Depletion
              Required to Account for
              the Observed Extinction
              }
\tablehead{
\colhead{Star}&
\colhead{$N_{\rm H}$}&
\colhead{$A_{V}$}&
\colhead{$R_{V}$}&
\colhead{$c_{1}^{\prime}$}&
\colhead{$c_{2}^{\prime}$}&
\colhead{$c_{3}^{\prime}$}&
\colhead{$c_{4}^{\prime}$}&
\colhead{$x_{0}$}&
\colhead{$\gamma$}&
\colhead{$A_{\rm int}^{a}$}&
\colhead{Proto-Sun$^{b}$}&
\colhead{}&
\colhead{}&
\colhead{B Stars$^{c}$}&
\colhead{}
\\
\cline{12-13}  
\cline{15-16}
\\
\colhead{}&
\colhead{($10^{21}\cm^{-2}\HH$)}&
\colhead{(mag)}&
\colhead{}&
\colhead{}&
\colhead{}&
\colhead{}&
\colhead{}&
\colhead{($\um^{-1}$)}&
\colhead{($\um^{-1}$)}&
\colhead{(mag$\cm^3\HH^{-1}$)}&
\colhead{$\VC/\rmH^{e}$}&
\colhead{$\cdust^{f}$}&
\colhead{}&
\colhead{$\VC/\rmH^{e}$}&
\colhead{$\cdust^{f}$}
}
\startdata

HD 027778 & 2.29$\pm$1.20 &  1.09$\pm$0.03 & 2.79$\pm$0.38 &  0.87$\pm$0.71& 0.32$\pm$0.01 & 1.31$\pm$0.10  & 0.24$\pm$0.01& 4.59$\pm$0.04 & 1.21$\pm$0.03&1.39E-25&1.69E-27&190&&2.05E-27&231\\
HD $037061^{d}$ & 5.37$\pm$1.23  &  2.40$\pm$0.21 & 4.29$\pm$0.21  &  1.54$\pm$0.00  & 0.00$\pm$0.00    & 0.31$\pm$0.00   & 0.05$\pm$0.00  & 4.57$\pm$0.00   & 0.90$\pm$0.00 &1.19E-25&1.19E-27&134&&1.55E-27&175\\
HD 037903 & 2.88$\pm$1.15 &  1.49$\pm$0.03 & 4.11$\pm$0.43 &  1.17$\pm$0.06  & 0.11$\pm$0.01  & 1.04$\pm$0.09  & 0.14$\pm$0.01   & 4.60$\pm$0.04  & 1.19$\pm$0.03&1.46E-25&1.86E-27&209&&2.22E-27&250 \\
HD 091824 & 1.44$\pm$1.12  &  0.60$\pm$0.06 & 3.10$\pm$0.31  &  0.74$\pm$0.41  &  0.29$\pm$0.07 &  1.19$\pm$0.35  & 0.14$\pm$0.02   &  4.62$\pm$0.09  & 1.00$\pm$0.10 &1.16E-25&1.11E-27&125&&1.47E-27&166  \\
HD $116852^{d}$ & 1.05$\pm$1.20 &  0.51$\pm$0.12 & 2.42$\pm$0.37  &  0.52$\pm$0.25 & 0.38$\pm$0.10 & 0.63$\pm$0.17  & 0.01$\pm$0.01 & 4.55$\pm$0.04 & 0.78$\pm$0.07 &1.13E-25&1.03E-27&117&&1.40E-27&158\\
HD 122879 & 2.19$\pm$1.26  &  1.41$\pm$0.04 & 3.17$\pm$0.20 &  1.15$\pm$0.08  & 0.15$\pm$0.02  &  0.53$\pm$0.06 & 0.09$\pm$0.01   & 4.57$\pm$0.04  & 0.74$\pm$0.02 &1.50E-25&1.96E-27&220&&2.32E-27&261\\
HD 147888 &  5.89$\pm$1.20 &  1.97$\pm$0.03 & 4.08$\pm$0.18 &  1.43$\pm$0.05  & 0.03$\pm$0.01  & 0.74$\pm$0.07  & 0.07$\pm$0.01   & 4.58$\pm$0.03  & 0.94$\pm$0.02&9.51E-26&5.93E-28&66.7&&9.54E-28&108 \\
HD 157857 & 2.75$\pm$1.17 &  1.37$\pm$0.05 & 3.15$\pm$0.30 &  0.78$\pm$0.09  & 0.22$\pm$0.02  & 1.47$\pm$0.12  & 0.11$\pm$0.01   & 4.54$\pm$0.03  & 0.96$\pm$0.02 &1.27E-25&1.39E-27&156&&1.75E-27&197\\
HD 185418 &  2.57$\pm$1.17 &  1.39$\pm$0.04 & 2.67$\pm$0.20 &  1.37$\pm$0.09  & 0.16$\pm$0.02  & 1.89$\pm$0.12  & 0.14$\pm$0.01   & 4.59$\pm$0.03  & 1.03$\pm$0.02 &1.55E-25&2.08E-27&234&&2.45E-27&275 \\
HD 192639 &  3.02$\pm$1.17 &  2.06$\pm$0.04 & 3.12$\pm$0.16 &  0.78$\pm$0.05  & 0.27$\pm$0.01   & 1.23$\pm$0.07   & 0.06$\pm$0.01   &  4.57$\pm$0.02 & 0.96$\pm$0.02 &1.80E-25&2.71E-27&304&&3.07E-27&345\\
HD 198781 & 1.41$\pm$1.15 &  1.03$\pm$0.03 & 2.77$\pm$0.29 &  0.44$\pm$0.09  & 0.42$\pm$0.02   & 2.09$\pm$0.23  & 0.13$\pm$0.02   & 4.69$\pm$0.06  & 1.36$\pm$0.03 &2.13E-25&3.53E-27&397&&3.89E-27&438\\
HD $203532^{d}$ & 2.75$\pm$1.17  &  0.94$\pm$0.11 & 3.37$\pm$0.24  &  0.76$\pm$0.16  & 0.27$\pm$0.03  & 1.50$\pm$0.25  & 0.20$\pm$0.04  & 4.59$\pm$0.01  & 1.26$\pm$0.04 &1.57E-25&2.13E-27&240&&2.50E-27 &281\\
HD 206773 &  1.78$\pm$1.12 &  2.16$\pm$0.04 & 4.08$\pm$0.29 &  0.84$\pm$0.07   &  0.18$\pm$0.01 & 0.75$\pm$0.07  & 0.04$\pm$0.01   & 4.57$\pm$0.04  & 1.04$\pm$0.02 &3.34E-25&6.52E-27&733&&6.88E-27&774\\
HD 207198 & 4.79$\pm$1.23  &  1.54$\pm$0.03 & 2.68$\pm$0.11 &  0.73$\pm$0.05  & 0.35$\pm$0.01   &  1.16$\pm$0.07 & 0.20$\pm$0.01    & 4.62$\pm$0.03  & 1.04$\pm$0.02 &9.30E-26&5.41E-28&60.8&&9.02E-28&102\\
HD 210809 & 2.14$\pm$1.20  &  0.98$\pm$0.05 & 3.10$\pm$0.31  &  0.60$\pm$0.31   & 0.32$\pm$0.02 & 0.83$\pm$0.08  & 0.15$\pm$0.01  & 4.60$\pm$0.09   & 1.00$\pm$0.10  &8.24E-26&2.77E-28&31.1&&6.38E-28&72.1 \\
HD $232522^{d}$ & 1.55$\pm$1.12 &  0.82$\pm$0.16 & 3.05$\pm$0.41  &  0.59$\pm$0.12 & 0.38$\pm$0.07 &  1.06$\pm$0.22  & 0.23$\pm$0.06  & 4.55$\pm$0.01  & 0.93$\pm$0.03 &1.54E-25&2.06E-27&232&&2.42E-27&273\\
\enddata 
\\
$^{a}$ The wavelength-integrated extinction  
       $\Aint \equiv\int_{912\Angstrom}^{10^{3}\um}
       A_\lambda/\NH\,d\lambda$\\
$^{b}$  The interstellar Si, Mg and Fe abundances 
        are taken to be that of proto-Sun (Lodders 2003)
        for which the total silicate volume per H atom 
        is $\Vsil/\rmH \approx 3.17\times10^{-27}\cm^{3}\HH^{-1}$.\\
$^{c}$  The interstellar Si, Mg and Fe abundances 
        are taken to be that of unevolved early 
        B stars (Przybilla et al.\ 2008)
        for which $\Vsil/\rmH \approx 2.52\times10^{-27}\cm^{3}\HH^{-1}$.\\
$^{d}$ Data taken from Valencic et al.\ (2004). 
       The others are taken from Gordon et al.\ (2009). \\ 
$^{e}$ The total volume of carbon dust per H atom
       (in unit of $\cm^3\HH^{-1}$) required to account 
       for the observed extinction.\\
$^{f}$ The carbon depletion $\cdust$
       (in unit of ppm) required to account 
       for the observed extinction.
\end{deluxetable}
%%%%%%%%%%%%%%   Table 1   %%%%%%%%%%%%%%

%%%%%%%%%%%%%%   Table 2   %%%%%%%%%%%%%%
\thispagestyle{empty}
\setlength{\voffset}{25mm}
\begin{deluxetable}{lccccccccccr}
\tablecolumns{10}
\tabletypesize{\scriptsize}
\tablewidth{0truein}
\center
\tablecaption{Model Parameters for Fitting
              the UV/Optical/Near-IR Extinction
              with A Mixture of Silicate/Graphite Dust
              \label{tab:modpara}
              }
\tablehead{
\colhead{Star}&
\colhead{$A_V/\NH$}&
\colhead{$\alpha_{\rm S}$}&
\colhead{$a_{c,{\rm S}}$}&
\colhead{$\alpha_{\rm C}$}&
\colhead{$a_{c,{\rm C}}$}&
\colhead{$\chi^{2}$}&
\colhead{$\cgas$}&
\colhead{$\cdust$}&
\colhead{$\sidust$}
\\
\colhead{}&
\colhead{($10^{-22} \magni\cm^{2}\,\rmH^{-1}$)}&
\colhead{}&
\colhead{($\um$)}&
\colhead{}&
\colhead{($\um$)}&
\colhead{}&
\colhead{(ppm)}&
\colhead{(ppm)}&
\colhead{(ppm)}
}
\startdata
HD 027778 &  4.76$\pm$2.50 &  -3.1& 0.14 & -3.0 & 0.10 &0.83 &  93$\pm$32  & 207  &  48.0   \\
HD 037061 &  4.47$\pm$1.10 & -2.0 & 0.17 & -2.4 & 0.05 &1.05 &  99$\pm$36  & 80.7 &  57.5   \\
HD 037903 &  5.17$\pm$2.07 & -2.7 & 0.23 & -3.4 & 0.20 &0.47 &  382$\pm$56 & 168  &  58.6   \\
HD 091824 &  4.17$\pm$3.27 & -3.1 & 0.24 & -3.8 & 0.25 &3.19 &  202$\pm$24 & 141  &  49.0   \\
HD 116852 &  4.86$\pm$5.67 & -2.7 & 0.08 & -3.4 & 0.15 &1.54 &  116$\pm$102& 177  &  43.0   \\
HD 122879 &  6.44$\pm$3.71 & -2.7 & 0.18 & -3.4 & 0.10 &6.19 &  365$\pm$94 & 184  &  64.0   \\
HD 147888 &  3.34$\pm$0.68 & -2.1 & 0.16 & -3.5 & 0.20 &2.28 &  98$\pm$23  & 101  &  35.0   \\
HD 157857 &  4.98$\pm$2.13 & -2.5 & 0.08 & -3.9 & 0.40 &0.14 &  112$\pm$19 & 225  &  31.0   \\
HD 185418 &  5.05$\pm$2.16 & -2.0 & 0.06 & -3.7 & 0.30 &4.14 &  173$\pm$34 & 310  &  36.0   \\
HD 192639 &  6.82$\pm$2.65 & -2.6 & 0.10 & -3.5 & 0.10 &3.16 &  138$\pm$24 & 227  &  60.8   \\
HD 198781 &  7.30$\pm$5.96 & -3.0 & 0.10 & -3.1 & 0.10 &0.12 &  193$\pm$31 & 313  &  72.9   \\
HD 203532 &  3.42$\pm$1.51 & -3.3 & 0.21 & -3.2 & 0.15 &0.87 &  85$\pm$28  & 165  &  28.8   \\
HD 206773 &  12.1$\pm$7.64 & -2.7 & 0.20 & -3.1 & 0.05 &3.40 &  464$\pm$57 & 206  &  143    \\
HD 207198 &  3.22$\pm$0.83 & -2.8 & 0.06 & -2.6 & 0.05 &3.49 &  69$\pm$21  & 168  &  21.0   \\
HD 210809 &  4.58$\pm$2.58 & -3.2 & 0.19 & -3.2 & 0.15 &4.12 &  321$\pm$70 & 118  &  28.0   \\
HD 232522 &  5.29$\pm$3.96 & -3.4 & 0.29 & -3.0 & 0.05 &4.12 &  317$\pm$46 & 183  &  63.9   \\
\enddata 
\\
\end{deluxetable}
%%%%%%%%%%%%%%   Table 2   %%%%%%%%%%%%%%

%%% Figure 12: gamma*c_3, gamma*c_4 vs C/H, Si/H %%%
%\begin{figure*}
%\centering
%\includegraphics[width=1.0\textwidth]{f12.eps}
%\caption{\footnotesize
%         \label{fig:gamma*c3_vs_C_Si_mod}
%         Correlation diagrams between the area of 
%	 2175\Angstrom \ bump ($\gamma c_3^{\prime}$) 
%         with the carbon depletion $\cdust$ (a) 
%         and silicon depletion $\sidust$ (b) 
%         derived from fitting the extinction 
%         of each sightline with a mixture of
%         silicate dust and graphite dust.
%         }
%\end{figure*}

\end{document}